\documentclass[12pt,a4paper]{article}
\usepackage{color, graphicx}

\def\slash#1{\mbox{$\not \!\! #1$}}
\def\Dslash{{\slash {\cal D}}}

\def\lvec#1{\setbox0=\hbox{$#1$}
    \setbox1=\hbox{$\scriptstyle\leftarrow$}
    #1\kern-\wd0\smash{
    \raise\ht0\hbox{$\raise1pt\hbox{$\scriptstyle\leftarrow$}$}}
    \kern-\wd1\kern\wd0}
\def\rvec#1{\setbox0=\hbox{$#1$}
    \setbox1=\hbox{$\scriptstyle\rightarrow$}
    #1\kern-\wd0\smash{
    \raise\ht0\hbox{$\raise1pt\hbox{$\scriptstyle\rightarrow$}$}}
    \kern-\wd1\kern\wd0}
\def\diracstar#1#2{
    \setbox0=\hbox{$\gamma$}\setbox1=\hbox{$\gamma_{#1}$}
    \gamma_{#1}\kern-\wd1\kern\wd0
    \smash{\raise4.5pt\hbox{$\scriptstyle#2$}}}
\def\tr{\,\hbox{tr}\,}

\newcommand{\beq}{\begin{equation}}
\newcommand{\eeq}{\end{equation}}
\newcommand{\beqn}{\begin{eqnarray}}
\newcommand{\eeqn}{\end{eqnarray}}
\newcommand{\nn}{\nonumber}

\usepackage{chir_layout}
\begin{document}
%%%%%%%%%%%%%%%%%%%%%%%%%
\begin{titlepage}
\pagestyle{empty}
\date{}
\title{
\bf A non-perturbative mechanism \\ for elementary particle mass generation 
\vspace*{3mm}}

\author{
        R.\ Frezzotti$^{a)}$\, and\, G.C.\ Rossi$^{a)\,b)}$
}
\maketitle
\begin{center}
  $^{a)}${\small Dipartimento di Fisica, Universit\`a di  Roma
  ``{\it Tor Vergata}'' \\ 
  and\\
  INFN, Sezione di Roma 2}\\
  {\small Via della Ricerca Scientifica, I-00133 Roma, Italy}\\
  $^{b)}${\small Centro Fermi - Museo Storico della Fisica}\\
  {\small Piazza del Viminale 1, I-00184 Roma, Italy}\\
\end{center}
%%%%%%%%%%%%%%%%%%%%%%%%%%%%%%%%%%%%%%%

\abstract{Taking inspiration from lattice QCD data, we argue that a finite non-perturbative  contribution to the quark mass is generated as a consequence of the dynamical phenomenon of spontaneous chiral symmetry breaking, in turn triggered by the explicitly breaking of chiral symmetry induced by the critical Wilson term in the action. In pure lattice QCD this mass term cannot be separated from the unavoidably associated linearly divergent contribution. However, if QCD is enlarged to a theory where also a scalar field is present, coupled to an SU(2) doublet of fermions via a Yukawa and a Wilson-like term, then in the phase where the scalar field takes a non-vanishing expectation value, a dynamically generated and ``naturally'' light fermion mass (numerically unrelated to the expectation value of the scalar field) is conjectured to emerge at a critical value of the Yukawa coupling where the symmetry of the model is maximally enhanced. Masses dynamically generated in this way display a natural hierarchy according to which the stronger is the strongest of the interactions the fermion is subjected to the larger is its mass.
}
\end{titlepage}
\newpage
     
\section{Introduction}
\label{sec:INTRO}

In this paper we argue that in lattice QCD with Wilson fermions~\cite{WIL} the dynamics of spontaneous chiral symmetry breaking (S$\chi$SB), in turn triggered by the explicit chiral breaking Wilson term in the action, is able to generate, even in the chiral limit, a non-perturbative (NP) finite (up to logs) mass contribution for the elementary fermions beneath the linearly divergent mass term that unavoidably goes with it.

If one can solve, as we are going to show in a simple renormalizable toy-model including QCD, the ``naturalness'' problem~\cite{THOOFT} associated to the need of ``fine tuning'' the parameters controlling the recovery of chiral symmetry in the critical theory, so as to be able to disentangle small (``finite''/non-perturbative) contributions from large (``infinite''/perturbative) terms, the ideas presented in this paper may open the way to a viable NP alternative to the Higgs mechanism for mass generation~\cite{FRNEW}. 

We shall argue that such non-perturbatively generated masses are proportional to the renormalization group invariant (RGI) scale, $\Lambda$, of the strong interactions which the fermions are subjected to. Effects of this kind are conjectured to stem from peculiar NP operator mixings that, though triggered by naively irrelevant Wilson-like terms in the action, survive the limit of infinite UV-cutoff. Quantitatively the resulting fermion mass terms of NP origin depend on the details of the UV-regularization of the model, thereby providing an example of {\em universality breaking} at the NP level~\footnote{A brief account of these ideas was presented at the LATTICE2013 Conference~\cite{RFPOS}.}. All these non-trivial expectations should be checked (or possibly falsified) by direct numerical simulations.

Interestingly the structure of the aforementioned enlarged toy-model is such that electro-weak interactions can be naturally introduced and mass terms for the weak gauge bosons are also generated by the same NP mechanism that is at work for the fundamental fermions~\cite{FRNEW}. 

Furthermore, if this toy model is extended by introducing in a gauge invariant way superstrongly interacting particles with RGI scale $\Lambda_{T}\gg \Lambda_{QCD}$, an interesting ordering of fermion masses emerges. In this situation, in fact, both quarks and superstrongly interacting fermions get a mass of the order of $\Lambda_{T}$ (the largest of the two RGI scales) but, as we shall see, scaled by powers of the strong ($g_s$) and superstrong ($g_T$) gauge coupling, respectively. Thus the difference in the strength of the two interactions is seen to be at the origin of the fact that the (top) quark mass is a fraction of the large scale $\Lambda_{T}$.  A crude phenomenological estimate gives for the superstrong scale a value in the few-TeV region if one has to get the NP generated top mass at the desired experimental value.

In a forthcoming paper~\cite{FRNEW} we will show that an extension of the model including, besides strong and superstrong forces, also electro-weak interactions and an appropriate set of fermion degrees of freedom to have gauge anomaly cancellation, can be elevated to a full beyond-the-Standard-Model model of elementary particles where all fermions (with the remarkable exception of neutrinos), as well as the weak bosons, acquire a mass proportional to $\Lambda_T$. Parametric mass hierarchy is a consequence of the fact that the non-perturbatively generated masses are scaled by powers of the coupling constants of the interactions the particle is subjected to. In particular weak gauge bosons and charged leptons masses are scaled by powers of the electro-weak gauge coupling constants. 

Moreover in models of this kind a host of interesting new phenomena arise, among which the presence in the spectrum of superstrongly bound states~\footnote{In the following, see sect.~\ref{sec:TECH}, we will suggestively term them ``techni-hadrons'', with an eye to the  bound states emerging in technicolor models~\cite{Weinberg:1979bn,Susskind:1978ms}, although our framework is very different from standard techni-color.} and gauge coupling unification at a very high O($10^{17}$)~GeV scale. 

The outline of the paper is as follows. In sect.~\ref{sec:LQCD} we discuss a NP mechanism that in lattice QCD (LQCD) with Wilson fermions~\cite{WIL} is capable to generate a finite (up to logs) term in the critical mass beneath the standard linearly divergent contribution. We provide both numerical evidence and theoretical arguments in favour of its existence. If we could single out this finite piece from beneath the linearly divergent term that goes with it, we could renormalize the theory in such a way that the NP finite term would play the r\^ole of a dynamically generated fermion mass. Within LQCD  such a ``fine tuning'' procedure is neither ``natural'' nor well defined. 

We show in sects.~\ref{sec:FINE1} and~\ref{sec:FINE2} how this ``naturalness problem''~\cite{THOOFT} can be circumvented in a model extension of QCD where a strongly interacting SU(2) doublet of fermions is coupled to a doublet of complex scalar fields via Yukawa and Wilson-like terms.  In sect.~\ref{sec:FINE1} we describe the symmetries of the model paying special attention to transformations of the chiral type and the associated Ward--Takahashi identities (WTIs). In sect.~\ref{sec:FINE2} we discuss how the physics of the model depends on the shape of the quartic scalar potential. If the latter has a single minimum (Wigner phase), we argue (subsect.~\ref{sec:WP}) that nothing special happens, in the sense that there is no trigger for the spontaneous breaking of chiral symmetry, hence no dynamical generation of fermion mass terms. But a critical value of the Yukawa coupling exists, at which the SU(2)$_L\times\,$SU(2)$_R$ fermion chiral transformations become a symmetry of the action, up to negligible (UV-cutoff)$^{-2}$ terms. In subsect.~\ref{sec:NGP} we discuss what happens in the much more interesting situation in which the scalar potential has the typical double-well shape (Nambu--Goldstone phase). In this case, at the same critical value of the Yukawa coupling (that was determined in the Wigner phase of the model), residual chiral breaking terms in the action (of the kind responsible for the similar phenomenon in LQCD at the critical mass, see sect.~\ref{sec:LQCD}) trigger the dynamical spontaneous breaking of the recovered chiral symmetry, yielding a NP finite (up to logs) mass to the fermions. In sect.~\ref{sec:TECH} we study the interesting situation occurring for fermion mass hierarchy if an extra family of fermions subjected to both strong and superstrong interactions is coupled to the model discussed in sects.~\ref{sec:FINE1} and~\ref{sec:FINE2}. Conclusions can be found in sect.~\ref{sec:CONOUT} together with a brief outlook on how ideas about the NP mass generation mechanism we propose can be extended to construct a complete beyond-the-Standard-Model model where fermion and weak boson mass hierarchy would ``naturally'' emerge.

\section{Inspiration and numerical evidence from lattice QCD}
\label{sec:LQCD}

As is well known, in LQCD with Wilson fermions~\cite{WIL} quark mass renormalization requires the subtraction of a linearly divergent counter-term, $m_{cr}\bar q q$ ($q $ being the $N_f$-flavour quark field), arising because the Wilson term in the lattice Lagrangian explicitly breaks chiral symmetry. In general $m_{cr}$ will have a formal small-$a$ expansion of the kind 
\begin{equation}
m_{cr} = \frac{c_0}{a}+c_1\Lambda_{QCD}+{\mbox{O}}(a)\, .
\label{CRM1}
\end{equation}
Eq.~(\ref{CRM1}) suggests that, if we could set the mass parameter, $m_0$, in the lattice fermion action just equal to the linearly divergent term $c_0/a$, a term proportional to $c_1 \Lambda_{QCD}$ would play the r\^ole of a quark mass in the renormalized chiral Ward--Takahashi Identities (WTIs) of the theory. 

To see how this can happen consider the renormalized axial (non-singlet) WTIs of lattice QCD. They read (in the notations of ref.~\cite{Bochicchio:1985xa}) 
\beq
\hspace{-.01cm}
\nabla_\mu\langle \hat J^f_{5\mu}(x) \hat O(0) \rangle \! = \!\langle  \Delta^f \hat O(0) \rangle \delta(x)\! +\! 2(m_0-\bar M(m_0))\langle P^f(x) \hat O(0) \rangle\!+\! {\mbox O}(a)\, ,\!\!\!
\label{AWTIRF}
\eeq
where $J^f_{5\mu}$, $f=1,2,\ldots N_f^2-1$, is the non-singlet axial current and $\bar M$ is the mixing coefficient between the axial variation of the Wilson term, $O_5^f$, and the pseudoscalar quark density, $P^f$. The hat denotes renormalized operators. In formulae we have
\beq
\hat O_5^f(x) = Z_5\Big{[} O_5^f(x) +\frac{2\bar M}{a}P^f(x)+\frac{Z_A-1}{a}\nabla_\mu J^f_{5\mu}(x)  \Big{]}\, ,\label{OFIVEN}
\eeq
where $\bar M(m_0)$ has the general expression 
\beq
\bar M (m_0)= \frac{c_0(1-d_1)}{a} + c_1(1-d_1)\Lambda_{QCD} + d_1 m_0+ {\mbox{O}}(a)\, ,
\label{MBAR}
\eeq
with the coefficients $c_0$, $c_1$ and $d_1$ functions of the gauge coupling, $g^2_s$. The coefficients $c_0$ and $d_1$ are present even in perturbation theory and their expansion starts at order $g_s^2$.

We recall that the solution of the equation $\bar M(m_0)=m_0$ is precisely $m_{cr}$ as given in eq.~(\ref{CRM1}). The key observation about eqs.~(\ref{AWTIRF})--(\ref{MBAR}) is that, if we could set 
\beq
m_0=\frac{c_0}{a}\ ,
\label{SUBTR}
\eeq  
the WTI~(\ref{AWTIRF}) would take the form 
\beq
\hspace{-.05cm}
\nabla_\mu\langle \hat J^f_{5\mu}(x) \hat O(0) \rangle \!= \!\langle  \Delta^f \hat O(0) \rangle \delta(x) - 2\, c_1(1-d_1)\Lambda_{QCD}\langle P^f(x) \hat O(0) \rangle + {\mbox O}(a)\, ,\!\!\label{FMASS}
\eeq
which shows that the quantity $-c_1(1-d_1)\Lambda_{QCD}$ plays the r\^ole of a non-per-turbatively  generated quark mass. In this situation, besides the standard perturbative quadratically divergent $c_0/a^2$ mixing between $O_5^f$ and $P_5^f$, one would have an extra NP contribution with a subleading linearly divergent $-2c_1(1-d_1)\Lambda_{QCD}/a$ coefficient. 

Notice that NP effects of this kind are immaterial for standard LQCD simulations, because $m_{cr}$ is always taken to be given by eq.~(\ref{CRM1}), i.e.\ as the value of $m_0$ at which the PCAC mass vanishes.

If one wants to make practical use of these considerations to construct a model where NP fermion mass generation takes place "naturally", one must be able to (positively) answer the following questions. 

1) Are there numerical indications for the existence of a term like the second one in the r.h.s.\ of~(\ref{CRM1}) in actual LQCD simulation data? 

2) Do we understand its possible dynamical origin? 

3) Are we in position to disentangle a (small) NP fermion mass from the much larger 
contribution that comes along with it when chiral symmetry is broken at a high momentum scale? 

\subsection{Some numerics}
\label{sec:NUM}

We start by examining the first among the three questions listed above and the one that has triggered this whole investigation. Hints for the existence of a non-vanishing $c_1 \Lambda_{QCD}$ term in eq.~(\ref{CRM1}) are numerically striking in Wilson LQCD simulations. 

Though the existence of this contribution may have been noticed in several simulations, its potential r\^ole for generating a genuine mass for the fermions was never taken in consideration because, as remarked above, a term of this kind (even if present) is anyway eliminated together with its linearly divergent counterpart, when the critical mass, determined by the vanishing of the PCAC mass, is subtracted out from the bare mass.  

\begin{figure}
\centerline{\includegraphics[scale=0.55,angle=-0]{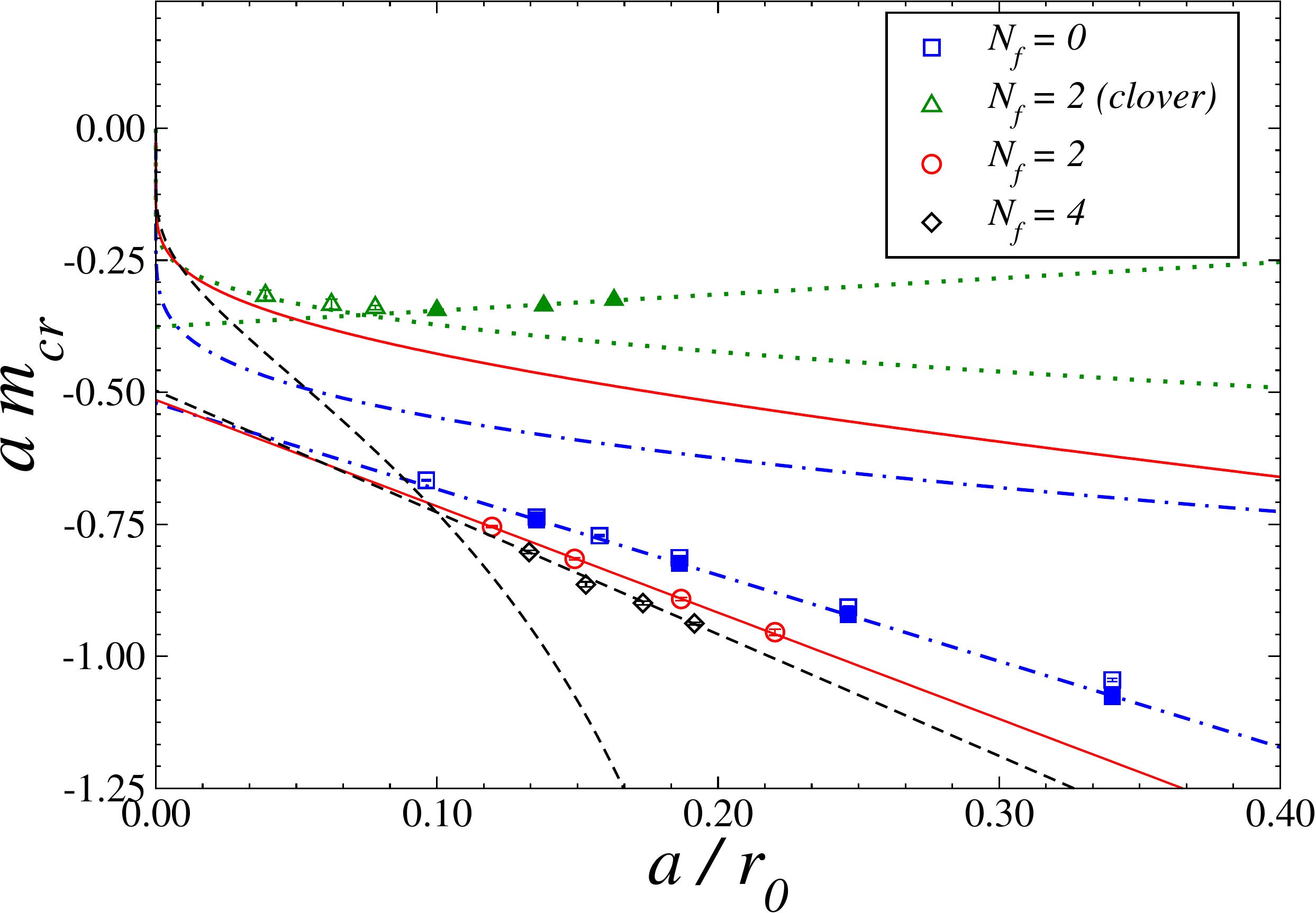}}
\caption{\small{The quantity $am_{cr}$ determined in Wilson LQCD simulations as a function of $a/r_0$. Black dots are the perturbative points of ref.~\cite{Skouroupathis:2008ry}. Blue squares are $N_f=0$ data from~\cite{Jansen:2005kk}. Red diamonds are $N_f=2$ data from~\cite{Baron:2009wt,Blossier:2010cr}. Black open circles are $N_f=4$ data from~\cite{Baron:2010bv,Baron:2011sf}. Green triangles are $N_f=2$ clover data from~\cite{DellaMorte:2004bc,Fritzsch:2012wq}. 
Straight lines denote the best fit (linear in $a/r_0$) of $am_{cr}$ simulation data.
In the ``$N_f=2$ (clover)'' case only the filled triangle data points are taken for the fit.
The four curves instead refer to 2-loop perturbative calculation of $am_{cr}$. 
Due to our trading of $g_0^2$ for $a/r_0$, the 2-loop curves bear, besides the 
intrinsic (main) error coming from truncation of the perturbative series,
a small uncertainty associated to the relation between $a/r_0$ and $g_0^2$.
This uncertainty can be converted into a relative error on $am_{cr}$ that
vanishes as $a/r_0 \to 0$ and in the region of simulation data amounts to 
about $2\%$, $2\%$, $3\%$ and $4\%$ for the $N_f=0$,
$N_f=2$ (clover), $N_f=2$ and $N_f=4$ curves, respectively.
For each lattice action the correspondence between $a/r_0$ and $g_0^2$
is established using the 2-loop formula
$a \, r_0^{-1}=(\Lambda_{\rm latt}r_0)^{-1} e^{-1/2 b_0 g_0^2} (b_0 g_0^2)^{-b_1/2b_0^2} \, ,$ where $b_{0,1}$ are the two universal coefficients of the $\beta$-function
and $\Lambda_{\rm latt}$ is evaluated by combining the exactly known ratio 
$ { \Lambda_{\rm \overline{MS}} }/{ \Lambda_{\rm latt } }$
with the determination (affected by errors at a few percents level) of 
$\Lambda_{\rm \overline{MS}} \,r_0$ from LQCD simulations and/or phenomenology.
}}
\label{fig:BAMBI}
\end{figure}

In fig.~\ref{fig:BAMBI} we report a compilation of perturbative and simulation data showing the behaviour of the value of $am_0$ at which $am_{PCAC}$ vanishes (that is to say the behaviour of $am_{cr}$), as a function of the dimensionless quantity $a/r_0$~\footnote{As customary, with $r_0$ we indicate the so-called Sommer parameter~\cite{Sommer:1993ce} that is used to scale dimensionful quantities in order to be able to meaningfully compare data obtained at different lattice spacings and/or in different LQCD formulations.}. Perturbative data are taken from the 2-loop calculations of ref.~\cite{Skouroupathis:2008ry} and plotted as function of $a/r_0$ after determining the relation between $g^2_0$ and $a\Lambda_{QCD}/r_0\Lambda_{QCD}$ combining  results from refs.~\cite{Luscher:1995nr,Martinelli:1998vt,Skouroupathis:2007mq,Carrasco:2014cwa,Agashe:2014kda}. Simulation data are extracted from measurements carried out in a number of LQCD studies employing Wilson fermions. We show four sets of data taken from refs.~\cite{Jansen:2005kk}, \cite{Baron:2009wt,Blossier:2010cr}, \cite{Baron:2010bv,Baron:2011sf} and~\cite{DellaMorte:2004bc,Fritzsch:2012wq}. 

Curves with black dashed, red full, blue dotted-dashed and green dotted points are the 2-loop perturbative estimates of $m_{cr}$ as function of $a/r_0$ for the four types of lattice actions for which non-perturbative values of the critical mass are also plotted.  Although perturbation theory can be considered to be reliable in a tiny range of values of $a/r_0$ (approximatively up to $a/r_0\simeq 0.01$), we have displayed the analytic behaviour of the perturbative curves throughout the whole span of the horizontal axis.

The three lower sets of points in fig.~\ref{fig:BAMBI} correspond to non-perturbative determinations of the critical mass performed at maximal twist using the Wilson twisted mass regularization of LQCD~\cite{Frezzotti:2000nk,Frezzotti:2003ni} in the quenched ($N_f=0$) approximation (blue squares~\cite{Jansen:2005kk}), with $N_f=2$ dynamical flavours (red diamonds~\cite{Baron:2009wt,Blossier:2010cr}) and with $N_f=4$ dynamical flavours (black open circles~\cite{Baron:2010bv,Baron:2011sf}), respectively. The green triangles correspond to the results obtained in~\cite{DellaMorte:2004bc,Fritzsch:2012wq} with $N_f=2$ dynamical flavours using clover-improved~\cite{Sheikholeslami:1985ij} Wilson fermions. 

In the present notations the slope of the fitted line through the non-perturbative data points of the figure is $c_1 \Lambda_{QCD}\times r_0$, i.e.\ the quantity of interest here. We see that the data of refs.~\cite{Jansen:2005kk}, \cite{Baron:2009wt,Blossier:2010cr} and~\cite{Baron:2010bv,Baron:2011sf} all exhibit a nice linear behaviour (with a mild $N_f$ dependence) in a wide window of $a/r_0$ values, which allows to ``identify'' a non-vanishing $c_1 \Lambda_{QCD}$.

A word of caution is in order here. On the one hand, strictly speaking there isn't any mathematically rigorous way to determine an $a/r_0$ range where one can consider  negligible both the logarithmic $a$-dependence of $am_{cr}$ governing its behaviour as $a\to 0$ (inherited from the behaviour of the renormalized gauge coupling), and the higher order lattice artefacts that become important at large enough $a$ values. 

On the other hand, the figure clearly shows that 1) the $a/r_0$ behaviour of the 2-loop perturbative curves is very different from that of the non-perturbative data, 2) it appears to be extremely difficult to provide a reasonable description of non-perturbative points without allowing for a linear term of the kind $c_1\Lambda_{QCD}a/r_0$ in $m_{cr}$.

Actually, as we said, a linear fit through the non-perturbative points of refs.~\cite{Jansen:2005kk}, \cite{Baron:2009wt,Blossier:2010cr} and~\cite{Baron:2010bv,Baron:2011sf} is quite good and gives for the numerical estimates of $c_1 \Lambda_{QCD}$ values around 700, 900 and 1000~MeV, respectively.

The Wilson clover improved data (green triangles) of refs.~\cite{DellaMorte:2004bc,Fritzsch:2012wq} are, instead, pretty flat implying that the $c_1$ coefficient is likely to be very small. This result is in line with our interpretation of the $m_{cr}$ behaviour as a function of $a$, according to which, as we shall argue in the next section, a non-zero slope is triggered by the chiral breaking terms in the Wilson action. The presence of the non-perturbatively tuned clover-term~\cite{Sheikholeslami:1985ij} in the lattice Lagrangian employed in refs.~\cite{DellaMorte:2004bc,Fritzsch:2012wq}, instead, effectively suppresses the relevant chiral breaking effects, thus leading to a much reduced value of the coefficient $c_1$ (O($a$) chiral breaking effects will be absent only in on-shell quantities).

The existence in $am_{cr}$ of NP O($a\Lambda_{QCD}$) corrections on top of the $c_0$ term should not come as a surprise. Indeed, there is an overwhelming evidence for similar cutoff effects in Wilson LQCD where they are seen to affect the correlation functions from which physical quantities like masses, operator matrix elements, etc., are extracted~\footnote{See e.g.~\cite{Gupta:1997nd} and~\cite{GL} for general arguments on the issue of non-perturbative O($a$) artefacts and ref.~\cite{Aoki:1999yr} for typical examples of this kind of effects on the hadron spectrum.}. On the other hand, it is known that in the absence of S$\chi$SB effects all (non-trivial) correlators of LQCD with massless Wilson fermions would be automatically O($a$) improved~\cite{Frezzotti:2003ni}, which is not the case.

We wish to conclude this section by observing that we expect a non-analytic dependence of $c_1$ on the Wilson $r$-parameter~\cite{WIL} as a footprint of the dynamical origin of the NP mass term $-c_1\Lambda_{QCD}$. Since the Wilson term is odd in $r$, $c_1$ should be proportional to $sign\,r$ (times an $r$-even coefficient). This behaviour is in analogy to what happens in QCD to the chiral condensate, $\langle \bar q q\rangle$, which (in the infinite volume limit) is proportional to $sign\,m_q$. Our point is that in both instances it is the dynamical breaking of chiral symmetry, triggered by either the (critical) Wilson term or by a non-zero mass term (or both), that is responsible for the occurrence of such NP dynamical phenomena. 

\subsection{The dynamical origin of the $c_1 \Lambda_{QCD}$ term}
\label{sec:DYNOR}

In this section we want to argue that in Wilson LQCD there is room for the appearance of a finite (up log's) contribution in $m_{cr}$, like the term $c_1 \Lambda_{QCD}$ we have introduced in eq.~(\ref{CRM1}) to fit simulation data. 

Two lines of reasoning can be followed. One is based on considerations stemming from the Symanzik expansion (subsect.~\ref{sec:SYMEXP}) and their implications for the lattice fermion self-energy (subsect.~\ref{sec:ENP}). The second relies on calculations directly performed in the basic lattice theory (subsect.~\ref{sec:SWDY}). Though none of the two can be rigorously pursued till the end (otherwise it would mean that we are in position of performing exact NP mass calculations in a regulazired field theory), the converging results provided by the two approaches make us confident that the numerical indication coming from the analysis of the data collected in fig.~\ref{fig:BAMBI} represents a real feature of $m_{cr}$.

\subsubsection{O($a\Lambda_{QCD}$) corrections: Symanzik expansion based argument}
\label{sec:SYMEXP}

In this subsection we want to provide arguments showing that the $c_1 \Lambda_{QCD}$ term emerges from a delicate interplay between O($a$) corrections to quark and gluon propagators and vertices ensuing from the spontaneous breaking of chiral symmetry, and the power-like divergence of the loop integration in self-energy diagrams where one Wilson term vertex is inserted. 

Indeed, peculiar NP O($a$) corrections, which are proportional to $\Lambda_{QCD}$ and independent of $m_0 - m_{cr}$, can be seen to affect lattice correlators. They can be geometrically described in terms of formal O($a$) contributions in the Symanzik expansion of lattice correlators~\cite{SYM}. The latter, in the limit $m_0 \to m_{cr}$ with $m_{cr}$ given by eq.~(\ref{CRM1}), can be expressed in the general form 
\begin{eqnarray}
\hspace*{-1.5cm} &&\langle O(x,x',...) \rangle\Big{|}^L\! = 
\langle O(x,x',...)  \rangle\Big{|}^C \!\!- 
a  \langle O(x,x',...) \!\int \! d^4 z L_5(z) \rangle\Big{|}^C\! + 
{\rm O}(a^2) \, ,\label{ALC1SYM}\\
\hspace*{-1.5cm} && O(x,x',...)  \Leftrightarrow  
A_\mu^b(x) A_\nu^c(x') \, , \;\;  q_{L/R}(x) \bar q_{L/R}(x') \, ,
\;\;  q_{L/R}(x) \bar q_{L/R}(x') A_\mu^b(y) \, , \label{ALC2SYM}
\end{eqnarray}
where $q = (q_1, \dots , q_{N_f})^T$ is a $N_f$-flavour quark field, $O$ is a (multi-)local, formally chiral invariant operator and $L_5$ is the $d=5$ chiral breaking Symanzik local effective Lagrangian (SLEL) operator, which in self-explanatory notations reads  
\beqn
L_5=b_{\sigma F} \, \bar q (i \sigma \cdot F) q + b_{DD} \, \bar q (-D \cdot D) q\, .\label{L51NOMASS}
\eeqn
The labels $|^C$ and $|^L$ are to remind that the correlators are evaluated in the massless limit of continuum and lattice QCD, respectively. 

The key remark about the expansion~(\ref{ALC1SYM}) is that the O($a$) continuum  correlators in the r.h.s.\ would vanish were it not for the phenomenon of S$\chi$SB, triggered by the chiral breaking (critical) Wilson term in the action. 

Symmetries and dimensional arguments allow us to determine the structure of the NP O($a$) contributions to quark and gluon propagators and $q\bar q$-gluon vertex in the expansion~(\ref{ALC1SYM}). The NP contributions we have identified add up to standard propagators and vertices, and for the operators listed in eq.~(\ref{ALC2SYM}) have the form 
\begin{eqnarray}
\hspace{-1.8cm}&&\Delta G_{\mu\nu}^{bc}(k) \Big{|}^L = - \alpha_s\, a\Lambda_{QCD} \,
\delta^{bc}\, \frac{ \Pi_{\mu\nu}(k) }{k^2}\,
 \,f_{AA}\Big{(}\frac{\Lambda_{QCD}^2}{k^2}\Big{)}  \, ,\label{DELTA1} \\
\hspace{-1.8cm}&&\Delta S_{LL/RR}(k) \Big{|}^L = - \alpha_s\,a\Lambda_{QCD} 
\,  \frac{i k_\mu  (\gamma_\mu)_{LL/RR} }{k^2} \,f_{q \bar q}\Big{(}\frac{\Lambda_{QCD}^2}{k^2}\Big{)} \, , \label{DELTA2}\\
\hspace{-1.8cm}&&\Delta \Gamma^{b,\mu}_{Aq\bar q}(k,\ell) \Big{|}^L \!\!
= \alpha_s\,a\Lambda_{QCD} \, ig_s \lambda^b \gamma_\mu f_{Aq \bar q}\Big{(}\frac{\Lambda_{QCD}^2}{k^2},\frac{\Lambda_{QCD}^2}{\ell^2},\frac{\Lambda_{QCD}^2}{(k+\ell)^2}\Big{)}\, ,
\label{DELTA3}
\end{eqnarray}
where $\Pi_{\mu\nu}(k)$ is the projector appropriate to the chosen gauge fixing condition. The O($a$) corrections displayed in eqs.~(\ref{DELTA1}), (\ref{DELTA2}) and~(\ref{DELTA3}) must be proportional to some non-vanishing power of $\alpha_s$, since in the free theory there would no such NP effect. One factor $\alpha_s$, indeed,  comes from the fact that the quark or gluon emitted from the $L_5$ vertex has to be absorbed somewhere in the diagram. That this power should be precisely equal to unit is a consequence of the structure of the Schwinger--Dyson equations for propagators and vertices (see e.g.\ fig.~4 of ref.~\cite{Alkofer:2000wg}, as well as refs.~\cite{Roberts:1994dr,Holl:2006ni} and Chapter 10 of the book~\cite{ITZUB} -- modulo the obvious modifications entailed here by the presence of the Wilson term in the action).  

In the formulae above we have left unspecified the scale at which the gauge coupling, $\alpha_s$, should be evaluated. The choice of this scale is not irrelevant as it will turn out to be a key feature to understand the details/numerics of the fermion mass hierarchy problem~\cite{FRNEW} (see the discussion in sect.~\ref{sec:TECH})). The occurrence of the RGI scale $\Lambda_{QCD}$ as a multiplicative factor in eqs.~(\ref{DELTA1}), (\ref{DELTA2}) and~(\ref{DELTA3}) signals the NP nature of the effect and appears to the first power for simple dimensional reasons. 

The scalar form factors $f_{AA}$, $f_{q\bar q}$ and $f_{Aq \bar q}$ are dimensionless functions depending on $\Lambda_{QCD}^2/({\mbox{momenta}})^2$ ratios. From the Symanzik  analysis of lattice artefacts, $a$-expansions like those in eqs.~(\ref{ALC1SYM}) are expected to be valid for squared momenta small compared to $a^{-2}$. Here we assume that the NP effects encoded in equations from~(\ref{DELTA1}) to~(\ref{DELTA3}) persist up to large (i.e.\ comparable to $a^{-1}$) momenta, and {\it conjecture} the asymptotic behaviour
\begin{eqnarray}
\hspace{-.8cm}&&f_{AA}\Big{(}\frac{\Lambda_{QCD}^2}{k^2}\Big{)} \stackrel{k^2\to\infty}\longrightarrow  {h_{AA}}\, , \,\,\,\,\, \label{ASYNT-FF-LQCD1} \\
\hspace{-.8cm}&&f_{q \bar q}\Big{(}\frac{\Lambda_{QCD}^2}{k^2}\Big{)} \stackrel{k^2\to\infty} \longrightarrow  h_{q\bar q}\, , \,\,\,\,\, \label{ASYNT-FF-LQCD2}  \\
\hspace{-.8cm}&&f_{Aq \bar q}\Big{(}\frac{\Lambda_{QCD}^2}{k^2},\frac{\Lambda_{QCD}^2}{\ell^2} \frac{\Lambda_{QCD}^2}{(k+\ell)^2}\Big{)}\stackrel{k^2,\ell^2,(k+\ell)^2\to\infty} \longrightarrow h_{q\bar q} \, ,\label{ASYNT-FF-LQCD3} 
\end{eqnarray}
where $h_{AA}$ and $h_{q \bar q}$ are O(1) constants and the last two limits are related by gauge invariance. 

It must be stressed that the asymptotic behaviour implied by eqs.~(\ref{DELTA2}) and~(\ref{ASYNT-FF-LQCD2}) is at variance with, and much softer than, the standard, large $k^2$ behaviour of the NP contributions to the quark propagator derived on the basis of the operator product expansion by Politzer~\cite{Politzer:1976tv} and Pascual-de Rafael~\cite{Pascual:1981jr} which would be unable to produce a term like $c_1\Lambda_{QCD}$ in the critical mass. The constant large momentum behaviour entailed by eqs.~(\ref{DELTA2}) and~(\ref{ASYNT-FF-LQCD2}) will be essential to generate a finite fermion mass contribution, as we are going to show below in this subsection.

\begin{figure}[htbp]
	\centering
\includegraphics[width=0.65\linewidth]{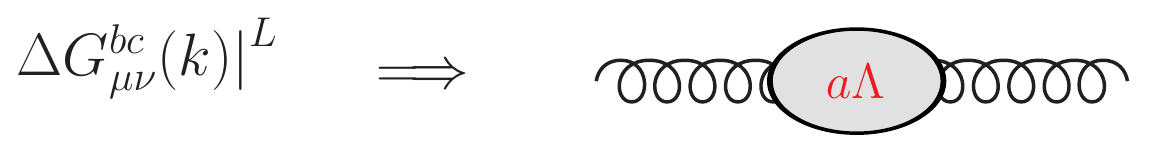}\\
		\vspace{.3cm}
\includegraphics[width=0.72\linewidth]{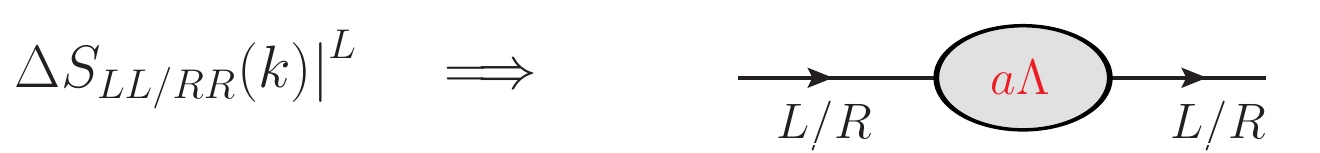}\\
	\vspace{.2cm}
\includegraphics[width=0.7\linewidth]{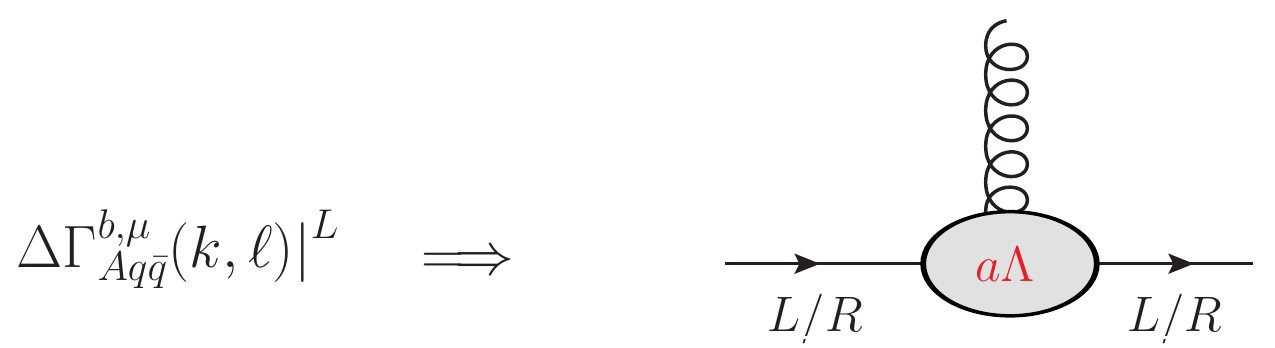}
\caption{\small{The NP O($a\Lambda_{QCD}$) terms contributing to the Symanzik expansion of quark and gluon propagators and $q\bar q$-gluon vertex (eqs.~(\ref{DELTA1}), (\ref{DELTA2}) and~(\ref{DELTA3})) are illustrated in the right panels. 
}}
\label{fig:FIG67}
\end{figure}

In the following we will represent the above O($a$) contributions to the quark and gluon propagator and to the quark--anti-quark--gluon vertex by the symbols 
shown in the right panels of fig.~\ref{fig:FIG67}~\footnote{Actually there are further NP corrections besides those displayed in eqs.~(\ref{DELTA1}), (\ref{DELTA2}) and~(\ref{DELTA3}). These are corrections to the Wilson term induced vertices and helicity-flipping quark propagator components. Based on LQCD symmetries, to leading order in $g_s^2$ (and $a$) a bookkeeping of all these NP effects can be obtained by constructing ``diagrams'' generated by the {\it ad hoc modified Feynman rules} that can be derived by adding to the LQCD Lagrangian the terms 
\beqn
\hspace{-.2cm}&&\Delta L \Big{|}_{ad\,hoc} = 
a\Lambda_{QCD} \alpha_s
\Big{\{}\frac{ h_{AA}}{4} (F \cdot F) + h_{q \bar q} (\bar q \slash D q) +
\nonumber \\\hspace{-.2cm}&& \hspace{1.0cm}+ 
h_{Wil}(-\frac{ar}{2}) (\bar q  D^2 q)
+ h_{Pau}(-\frac{ar}{2}) (\bar q  i\sigma \cdot F q) \Big{\}} \, .\nn
\eeqn
In order to avoid any misunderstanding or confusion it is important to stress that the augmented Lagrangian, $L_{LQCD}+\Delta L |_{ad\,hoc}$, can only be used to gain  insights on the structure of possible NP effects in a sort of heuristic mixed approach where NP effects are incorporated in an otherwise perturbative calculation (like in fig.~\ref{fig:FIG6}). In other words the form of $\Delta L |_{ad\,hoc}$ is such so as to reproduce (to leading order in $g^2_s$) the O($a$) results of the Symanzik expansion, with the inclusion of NP corrections. Of course the full and complete computation, from which all the NP effects we have described above are expected to emerge, should be carried out by using the fundamental LQCD Lagrangian.}.

\subsubsection{The emergence of a NP quark mass contribution}
\label{sec:ENP}

To explicitly see how finite $c_1\Lambda_{QCD}$ term can arise in $m_{cr}$ let us consider the $L-R$ component of the quark propagator and look for possible O($a^0\Lambda_{QCD}$) NP mass-like contribution in LQCD. The precise value of the renormalized quark mass is unimportant here. 

\begin{figure}[htbp]
\centerline{\includegraphics[scale=0.40,angle=0]{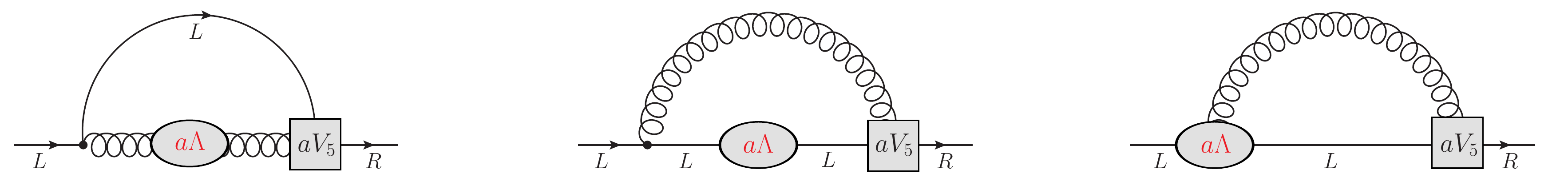}}
\caption{\small{Typical lowest order lattice ``diagrams'' giving rise to dynamically generated quark mass terms ($L$ and $R$ are quark-helicity labels). The square box represents the Wilson vertex and the grey blob the non-perturbative $a\Lambda_{QCD}\alpha_s$ effect stemming from the second term in the r.h.s.\ of eqs.~(\ref{ALC1SYM}).}}
\label{fig:FIG6}
\end{figure}

Finite (up to logs) dynamical mass terms get generated from ``diagrams'' like the typical ones shown in fig.~\ref{fig:FIG6} provided the lattice propagators and vertices receive the NP O($a \Lambda_{QCD}$) corrections of eqs.~(\ref{DELTA1}), (\ref{DELTA2}) and~(\ref{DELTA3})) (that are graphically summarised in Fig.~\ref{fig:FIG67}). 

We note that the insertion of these O($a \Lambda_{QCD}$) corrections in our ideal NP evaluation of the ($L-R$ component of) the quark propagator can be justified on the basis of the exact Schwinger--Dyson equation for the quark propagator, the structure of which, in the simpler case of continuum QCD, is discussed e.g.\ in ref.~\cite{Alkofer:2000wg} (see also fig.~4 there).

Let us consider as an example the case of the diagram in the central panel of fig.~\ref{fig:FIG6}. In the $a\to 0$ limit, the loop momentum (call it $k$) counting gives (for small external momentum) factors $a\Lambda_{QCD}\alpha_s k_\mu/k^2$ and $1/k^2$ from the NP contribution to the quark propagator and the standard gluon propagator, respectively, and a factor $a k_\mu$ from the derivative coupling of the Wilson vertex. If we assume the constant asymptotic behaviour~(\ref{ASYNT-FF-LQCD2}), we recognise that the multiplicative $a^2$ power is exactly compensated by the quadratic divergency of the loop-integral. Including an $\alpha_s$ factor from the gluon loop, one thus gets schematically a fermion mass term of the order
\begin{equation}
a\Lambda_{QCD}g^2_s\alpha_s  \int^{1/a} d^4 k \frac{k_\mu}{k^2}\frac{1}{k^2} ak_\mu \sim g^2_s\alpha_s \Lambda_{QCD}\, .
\label{CONS}
\end{equation}
Other ``diagrams'' give similar NP mass contributions yielding in eq.~(\ref{CRM1}), as well as in eq.~(\ref{MBAR}), to lowest order in the gauge coupling, the result $c_1 \sim {\mbox{O}}(\alpha_s^2)$. 

Summarizing, the argument shows that relative O($a\Lambda_{QCD} \alpha_s$) corrections to propagators and vertices have the potential of generating NP O($\alpha_s^2\Lambda_{QCD}$) corrections to the quark self-energy.

\subsubsection{Argument based on the spectral Dirac operator density}
\label{sec:SWDY} 

A second line of arguments one can give to support the emergence of a finite 
quark mass term of dynamical origin is based on the occurrence in the spectral
density of the Wilson Dirac operator of NP contributions $\propto \Lambda_{QCD}$
that are related to the phenomenon of spontaneous chiral symmetry breaking. In this approach NP chiral breaking effects are incorporated in the quark propagator by assuming that the (gluon averaged) eigenvalue density of the Wilson--Dirac operator admits an expansion of the type 
\beq
\widehat \rho_{D}(\lambda)=r_1\Lambda_{QCD}^3+r_2\Lambda_{QCD}^2|\lambda|+r_3\Lambda_{QCD}|\lambda|^2+r_4|\lambda|^3+ \dots\, . 
\label{RHOAV}
\eeq
The first and the last term are well-known and correspond to the Banks--Casher~\cite{Banks:1979yr} and the perturbative contribution, respectively. Theoretical arguments in favour of the existence of the second and third term are given in refs.~\cite{Smilga:1993in,Damgaard:1998xy,Osborn:1998qb}. Numerical indication for deviations from the purely Casher--Banks plus perturbative behaviour can be found in refs.~\cite{Cichy:2013gja,Cichy:2013eoa}

The evaluation of the quark self-energy in the fundamental theory is quite complicated as, in the spirit (again) of the Schwinger--Dyson equations, it requires first to compute the relevant NP corrections to the gluon and quark propagator and to the quark--anti-quark--gluon vertex and then to insert these building blocks in higher order self-energy diagrams. 

In Appendix~A we present a prototype calculation of the quark self-mass which indeed indicates that the NP $\Lambda_{QCD}|\lambda|^2$ term in the Dirac--Wilson eigenvalue density generates the sought for $c_1 \Lambda_{QCD}$ finite (up to log's) contribution to the quark mass term.  This analysis has also the merit of showing that the constant asymptotic behaviour of the NP correction terms displayed in eqs.~(\ref{ASYNT-FF-LQCD1}), (\ref{ASYNT-FF-LQCD2}) and~(\ref{ASYNT-FF-LQCD3}) is the correct one, or in other words that the NP contributions to effective propagators and vertices of eqs.~(\ref{DELTA1}), (\ref{DELTA2}) and~(\ref{DELTA3}), which were argued to occur on the basis of a Symanzik expansion of LQCD correlators, do persist up to momenta of the order of the UV cutoff.

\subsection{Is the mass subtraction~(\ref{SUBTR}) well defined and ``natural''?}
\label{sec:SUBNAT}
 
At this point the key question is whether it is sensible, i.e.\ well defined and ``natural", to adopt the mass renormalization prescription specified in eq.~(\ref{SUBTR}), or in other words, whether it is possible to subtract out from $m_0$ just the $c_0/a$ counter-term and not the whole critical mass that is obtained from the condition of ``vanishing PCAC mass'' (i.e.\ restoration of non-singlet axial WTI's). We must stress that the nature of this problem is conceptually the same as that of the naturalness problem~\cite{THOOFT} in the SM. 

Within LQCD with Wilson fermions the answer to the question above is negative, i.e.\ no solution exists to this naturalness problem, essentially because in this theory there is only one operator, namely $P^f$, of dimension three which $O_5^f$ can mix with. As a consequence no symmetry-based criterion can be found allowing to single out a finite term from beneath a linearly diverging one in the mixing coefficient between $X^f = aO_5^f$ and $P^f$.

We will show in the next section that an extension of QCD, where a doublet of strongly interacting fermions is coupled to a doublet of complex scalar fields via Yukawa and Wilson-like terms, provides a framework in which the fine tuning problem (or better the appropriate analog of the quark mass subtraction~(\ref{SUBTR}) in Wilson LQCD) appears to have a ``natural'' solution if one requires that the renormalized theory enjoys an enlarged fermionic symmetry of the chiral type. In the phase where the scalar field acquires a vacuum expectation value (vev), this symmetry turns out to be dynamically broken by a NP mechanism analogous to the one ultimately responsible for the generation of the $c_1 \Lambda_{QCD}$ term in LQCD.  

The key difference with LQCD is that in this extended theory a new, genuinely NP operator of dimension three appears in the renormalized WTIs. This purely NP operator is seen to 
be multiplied by a well defined and ``naturally" light effective fermion mass of dynamical origin, that interestingly is proportional to $\Lambda_{QCD}$ and independent of the scalar field vev.

\section{Light mass fermions with natural fine tuning: a toy model}
\label{sec:FINE1}

If we want to employ a NP mechanism of the kind outlined in sect.~\ref{sec:LQCD} for fermion mass generation, we have to provide a (good) reason for choices like $m_0=c_0/a$ in eq.~(\ref{SUBTR}), or more generally of special values for chiral-restoring counter-term parameters, so as to avoid an undesirable ``fine tuning" problem. From the arguments developed in sect.~\ref{sec:SUBNAT} it should be clear that such a reason must necessarily lie outside the LQCD theory we have considered up to now and must be based on symmetry and renormalizability considerations. 

In this and next section we present a concrete example of a possible theoretical scheme where a ``light" fermion mass term can be dynamically generated with no ``unnatural'' fine tuning~\cite{THOOFT}.

\subsection{Coupling fermions to non-Abelian gauge fields and scalars}
\label{sec:FINE1-A}

Let us consider a toy-model described by the formal Lagrangian 
\beqn
\hspace{-1.6cm}&&{\cal L}_{\rm{toy}}(Q,A,\Phi)= {\cal L}_{kin}(Q,A,\Phi)+{\cal V}(\Phi)
+{\cal L}_{Wil}(Q,A,\Phi) + {\cal L}_{Yuk}(Q,\Phi) \, ,\label{SULL}
\eeqn
\beqn
\hspace{-1.6cm}&&{\cal L}_{kin}(Q,A,\Phi)= \frac{1}{4}(F\cdot F)+\bar Q_L\Dslash Q_L+\bar Q_R\Dslash \,Q_R+\frac{1}{2}{\tr}\big{[}\partial_\mu\Phi^\dagger\partial_\mu\Phi\big{]}\label{LKIN}\\
\hspace{-1.6cm}&&{\cal V}(\Phi)= \frac{\mu_0^2}{2}{\tr}\big{[}\Phi^\dagger\Phi\big{]}+\frac{\lambda_0}{4}\big{(}{\tr}\big{[}\Phi^\dagger\Phi\big{]}\big{)}^2\label{LPHI}\\
\hspace{-1.6cm}&&{\cal L}_{Wil}(Q,A,\Phi)= \frac{b^2}{2}\rho\,\big{(}\bar Q_L{\overleftarrow{\cal D}}_\mu\Phi {\cal D}_\mu Q_R+\bar Q_R \overleftarrow{\cal D}_\mu \Phi^\dagger {\cal D}_\mu Q_L\big{)}
\label{LWIL} \\
\hspace{-1.6cm}&&{\cal L}_{Yuk}(Q,\Phi)=\
  \eta\,\big{(} \bar Q_L\Phi Q_R+\bar Q_R \Phi^\dagger Q_L\big{)}
\label{LYUK} \, ,
\eeqn 
where $b^{-1}=\Lambda_{UV}$ is the UV-cutoff~\footnote{To avoid confusion in this 
and the following sections the UV-regularization scale will be denoted by $b^{-1}$.}. The parameter $\rho$ in eq.~(\ref{LWIL}) is of no relevance for the naturalness arguments we are going to develop in this paper. It has been, however, already introduced here as a preparation because the tuning of $\rho$ will be instrumental for solving the naturalness problem when electro-weak interactions are present~\cite{FRNEW}. 

Apart from the cutoff scale, the details of UV-regularization are left unspecified here as they will be immaterial for the following qualitative discussion which is mainly based on symmetry considerations. Remarks on the impact of the UV-regularization details (universality violations) on the actual {\em magnitude} of the NP fermion masses that may be dynamically generated can be found in sects.~\ref{sec:THEOREMA} and~\ref{sec:TECH}. 

The Lagrangian~(\ref{SULL}) describes a non-Abelian gauge model where an SU(2) doublet of strongly interacting fermions is coupled to a complex scalar field via Wilson-like (eq.~(\ref{LWIL})) and Yukawa (eq.~(\ref{LYUK})) terms.

For short we have used a compact SU(2)-like notation where $Q_L=(u_L\,\,d_L)^T$ and $Q_R=(u_R\,\,d_R)^T$ are fermion iso-doublets and $\Phi$ is a $2\times2$ matrix with $\Phi=(\phi,-i\tau^2 \phi^*)$ and $\phi$ an iso-doublet of complex scalar fields. We immediately notice that this structure is ready to be gauged to accommodate electro-weak interactions~\cite{FRNEW}.

The term ${\cal V}(\Phi)$ in eq.~(\ref{LPHI}) is the standard quartic scalar potential where the (bare) parameters $\lambda_0$ and $\mu_0^2$ control the self-interaction and the mass of the scalar field. In the equations above we have introduced the covariant derivatives
\beq
{\cal D}_\mu=\partial_\mu -ig_s \lambda^a A_\mu^a \, , \qquad
\overleftarrow{\cal D}_\mu =\overleftarrow{\partial}_\mu +ig_s \lambda^a A_\mu^a \, ,\label{COVG}
\eeq
where $A_\mu^a$ is the gluon field ($a=1,2,\dots, N_c^2-1$) with field strength $F_{\mu\nu}^{a}$. A crucial r\^ole in the model is played by the $d=4$ Yukawa term ${\cal L}_{Yuk}$ and the $d=6$ operator ${\cal L}_{Wil}$. By dimensional reasons the latter enters the Lagrangian multiplied by $b^2$. We have denoted it with the subscript ``${Wil}$", because, as far as symmetries of chiral type are concerned, it will play a r\^ole similar to that of the Wilson term in standard Wilson LQCD~\footnote{Actually for this purpose also other operators, like e.g.\ $\bar Q_L\Phi \,i (\sigma \cdot F) Q_R + \bar Q_R \Phi^\dagger \,i(\sigma \cdot F) Q_L$, would equally well do the job. Lagrangian terms with $d=6$ are part of the UV-regularization of the model, which is not fully specified at this stage. Anyway, in our approach at least some $d \geq 6$ operator that breaks purely fermionic chiral symmetries must be assumed to occur in the UV-regulated model.}. 

\subsection{Symmetries of the models in the ${\cal L}_{\rm{toy}}$ class }
\label{sec:GSM}

Besides the obvious Lorentz, gauge and $C$, $P$, $T$ symmetries (see Appendix~B), ${\cal L}_{\rm toy}$ is invariant under the following (global) transformations 
\beqn
\hspace{-2.cm}&&\bullet\,\chi_L:\quad \tilde\chi_L\otimes (\Phi\to\Omega_L\Phi) 
\label{CHIL}\\
\hspace{-2.cm}{\mbox{where}}\nn\\
\hspace{-2.cm}&&\tilde\chi_L : \left \{\begin{array}{l}     
Q_L\rightarrow\Omega_L Q_L  \\
\hspace{4cm}\Omega_L\in {\mbox{SU}}(2)_L\\
\bar Q_L\rightarrow \bar Q_L\Omega_L^\dagger \\ 
\end{array}\right . \label{GTWT}
\eeqn
\beqn
\hspace{-2.cm}&&\bullet\,\chi_R:\quad \tilde\chi_R\otimes (\Phi\to\Phi\Omega_R^\dagger) 
\label{CHIR}\\
\hspace{-2.cm}{\mbox{where}}\nn\\
\hspace{-2.cm}&&\tilde\chi_R : \left \{\begin{array}{l}     
Q_R\rightarrow\Omega_R Q_R \\
\hspace{4.cm}\Omega_R\in {\mbox{SU}}(2)_R \\
\bar Q_R\rightarrow \bar Q_R\Omega_R^\dagger \\ 
\end{array}\right . \label{GTCT}
\eeqn
The conserved currents corresponding to the exact $\chi_L \times\chi_R$ symmetry read ($i=1,2,3$)
\beqn
\hspace{-.8cm}&&J_\mu^{L\,i}= \bar Q_L\gamma_\mu\frac{\tau^i}{2}Q_L-\frac{1}{2}{\mbox{\rm tr}}\big{[}\Phi^\dagger\frac{\tau^i}{2}\partial_\mu\Phi-(\partial_\mu\Phi^\dagger)\frac{\tau^i}{2}\Phi\big{]}+\nn\\
\hspace{-.8cm}&&\phantom{J_\mu^{W\,i}}-\frac{b^2}{2}\rho\,\big{(} \bar Q_L \frac{\tau^i}{2}\Phi {\cal D}_\mu Q_R-\bar Q_R \overleftarrow{\cal D}_\mu \Phi^\dagger \frac{\tau^i}{2} Q_L\big{)}\, , \label{CCL}\\
\hspace{-.8cm}&&J_\mu^{R\,i}= \bar Q_R\gamma_\mu\frac{\tau^i}{2}Q_R-\frac{1}{2}{\mbox{\rm tr}}\big{[}(\partial_\mu\Phi^\dagger)\Phi\frac{\tau^i}{2}-\frac{\tau^i}{2}\Phi^\dagger (\partial_\mu\Phi)\big{]}+\nn\\
\hspace{-.8cm}&&\phantom{J_\mu^{W\,i}}-\frac{b^2}{2}\rho\,\big{(}\bar Q_R \frac{\tau^i}{2}\Phi^\dagger {\cal D}_\mu Q_L-\bar Q_L \overleftarrow{\cal D}_\mu \Phi \frac{\tau^i}{2} Q_R\big{)}\, ,
\label{CCR}
\eeqn
giving rise to the WTIs
\beqn
&&\partial_\mu \langle J^{L\, i}_\mu(x) \,\hat O(0)\rangle = 
\langle \Delta_{L}^i \hat O(0)\rangle\delta(x) \, ,
\label{CHIL-WTI} \\
&&\partial_\mu \langle J^{R\, i}_\mu(x) \,\hat O(0)\rangle = 
\langle \Delta_{R}^i \hat O(0)\rangle\delta(x) \, ,
\label{CHIR-WTI}
\eeqn
where $\hat O$ is a renormalized (multi-)local operator and $\Delta_{L}^i\hat O$ and $\Delta_{R}^i\hat O$ are the variations of $\hat O$ under $\chi_L$ and $\chi_R$, respectively. 

The model~(\ref{SULL}) is power-counting renormalizable (as LQCD is) with counter-terms constrained by the exact symmetries of the Lagrangian. We note in particular that, owing to the presence of the scalar field and the related exact $\chi_L \times \chi_R$ symmetry,
no power divergent fermion mass terms can be generated in perturbation theory. 

For later use we remark that the renormalized correlation functions of the model~(\ref{SULL}) admit a small-$b$ Symanzik-like expansion where only cutoff corrections with {\em even} powers of $b$ appear. The absence of odd powers relies on the invariance of the Lagrangian~(\ref{SULL}) under the discrete transformation, ${\cal D}_d$, that consists in multiplying each field by the factor $e^{i\pi d} = (-1)^d$, with $d$ its naive dimension, and simultaneously changing sign to its space-time argument~\cite{Frezzotti:2003ni}~\footnote{${\cal D}_d$ can also be viewed as the product of parity, time reversal and the discrete chiral transformations ${\cal R}_5\times U(1)_F(\pi/2)$, where ${\cal R}_5\equiv V^1_0(\pi/2)V^2_0(\pi/2)A_0^3(\pi/2)$ is a product of three discrete (non-singlet) chiral transformations and $U(1)_F(\pi/2)$ is a discrete transformation (the one under which $Q \to i Q$, $\bar{Q} \to -i \bar{Q}$) of the global symmetry group $U(1)_F$ corresponding to fermion number conservation. Although S$\chi$SB can affect the way the ${\cal R}_5$ symmetry is realized, this symmetry still constrains the operators entering the SLEL to only the even-dimensional ones.}. 
One checks that only operators with even (naive) dimension can occur in the formal SLEL that generates the small-$b$ expansion of correlators.

\subsection{Bare WTIs of $\tilde\chi_L \times \tilde\chi_R$ transformations}
\label{sec:chitilde}

For generic values of the parameters, ${\cal L}_{\rm{toy}}$ is not invariant under the chiral transformations $\tilde\chi_L$ (eq.~(\ref{GTWT})) and $\tilde\chi_R$ (eq.~(\ref{GTCT})) that leave the scalar field untouched. Rather these transformations give rise to the (bare) WTIs 
\beqn
\hspace{-0.7cm}&&\partial_\mu \langle \tilde J^{L\, i}_\mu(x) \,\hat O(0)\rangle = \langle \tilde\Delta_{L}^i \hat O(0)\rangle\delta(x) -
\eta \,\langle \big{(} \bar Q_L\frac{\tau^i}{2}\Phi Q_R-\bar Q_R\Phi^\dagger\frac{\tau^i}{2}Q_L \big{)}(x)\,\hat O(0)\rangle \!+\nn\\
\hspace{-1.2cm}&&\phantom{\partial_\mu J^{L\, i}_\mu}-\frac{b^2}{2}\rho\,\langle\Big{(} \bar Q_L\overleftarrow {\cal D}_\mu\frac{\tau^i}{2}\Phi{\cal D}_\mu Q_R-\bar Q_R\overleftarrow {\cal D}_\mu \Phi^\dagger\frac{\tau^i}{2}{\cal D}_\mu Q_L\Big{)}(x)\,\hat O(0)\rangle\, ,\label{CTLTI} 
\eeqn
\beqn
\hspace{-0.7cm}&&\partial_\mu \langle \tilde J^{R\, i}_\mu(x) \,\hat O(0)\rangle = \langle \tilde\Delta_{R}^i \hat O(0)\rangle\delta(x) -
\eta \,\langle \big{(} \bar Q_R\frac{\tau^i}{2}\Phi^\dagger Q_L-\bar Q_L\Phi\frac{\tau^i}{2} Q_R \big{)}(x)\,\hat O(0)\rangle \!+\nn\\
\hspace{-1.2cm}&&\phantom{\partial_\mu J^{L\, i}_\mu}-\frac{b^2}{2}\rho\,\langle\Big{(} \bar Q_R\overleftarrow {\cal D}_\mu\frac{\tau^i}{2}\Phi^\dagger{\cal D}_\mu Q_L-\bar Q_L\overleftarrow {\cal D}_\mu \Phi\frac{\tau^i}{2}{\cal D}_\mu Q_R\Big{)}(x)\,\hat O(0)\rangle\, ,\label{CTRTI} 
\eeqn
where $\tilde\Delta_{L}^i\hat O$ and $\tilde\Delta_{R}^i\hat O$ are the variations of $\hat O$ under $\tilde\chi_L$ and $\tilde\chi_R$, respectively. The non-conserved currents associated to the transformations $\tilde\chi_L$ and $\tilde\chi_R$ are 
\beq
\tilde J_\mu^{L\,i}= \bar Q_L\gamma_\mu\frac{\tau^i}{2}Q_L -\frac{b^2}{2}\rho\Big{(}\bar Q_L\frac{\tau^i}{2}\Phi {\cal D}_\mu Q_R - \bar Q_R\overleftarrow {\cal D}_\mu\Phi^\dagger\frac{\tau^i}{2} Q_L\Big{)}\, ,
\label{JCLT}
\eeq
\beq
\tilde J_\mu^{R\,i}=\bar Q_R\gamma_\mu\frac{\tau^i}{2}Q_R -\frac{b^2}{2}\rho\Big{(}\bar Q_R\frac{\tau^i}{2}\Phi^\dagger {\cal D}_\mu Q_L - \bar Q_L\overleftarrow {\cal D}_\mu\Phi\frac{\tau^i}{2} Q_R\Big{)}\, ,
\label{JCRT}
\eeq
and differ from the conserved ones, $J^{L\, i}_\mu$ and $J^{R\, i}_\mu$, only because in the latter a contribution bilinear in the scalar field coming from the $\Phi$-kinetic term appears. 

At this stage, owing to the freedom in choosing the parameter $\eta$ (and $\rho$), we have a family of models endowed with exact $\chi_L \times \chi_R$ invariance, but where in general the transformations $\tilde\chi_L$ and $\tilde\chi_R$ are not symmetries of ${\cal L}_{\rm toy}$. In the following we will show that there exists a ``critical'' value of the Yukawa coupling, $\eta_{cr}(g_s^2,\rho,\lambda_0)$, at which, up to negligibly small O($b^2$) cutoff effects, the chiral $\tilde\chi_{L}\times\tilde\chi_{R}$-transformations are elevated to symmetries of the theory. This property can be regarded as an extension of the Golterman--Petcher symmetry~\cite{Golterman:1989df} valid for the Higgs-Yukawa model to the present case where fermions interact also with gauge fields.

Symmetry restoration does not depend on the fine details of the UV-regularization of the model~(\ref{SULL}), which in fact has not been fully specified, except for the crucial inclusion of a $\tilde\chi_L \times \tilde\chi_R$-breaking Wilson-like $d=6$ term and the Yukawa terms that unavoidably goes with it. Upon changing the UV-regularization details while preserving the exact symmetries of ${\cal L}_{\rm toy}$, 
no new Lagrangian terms with $d \leq 4$ can be generated via loop corrections, implying that just the numerical value of $\eta_{cr}$ and of other bare parameters will be affected. 

To give a precise meaning to the criterion of ``$\tilde\chi_{L}\times\tilde\chi_{R}$-symmetry enhancement" we need to study the mixing pattern of the operators appearing in the r.h.s.\ of the WTIs~(\ref{CTLTI}) and~(\ref{CTRTI}) and proceed to renormalization. 

\subsection{Renormalizing $\tilde{\chi}_{L} \times \tilde{\chi}_{R}$ WTIs}
\label{sec:MSE-PT}

As we just said, in order to renormalize the WTIs~(\ref{CTLTI}) and~(\ref{CTRTI}) we have to work out the mixing pattern of the $d = 6$ operators 
\beqn
\hspace{-.5cm}&& O_{6}^{L\, i} = \frac{1}{2} \rho
\Big{[}\bar Q_L\overleftarrow {\cal D}_\mu\frac{\tau^i}{2}\Phi{\cal D}_\mu Q_R
-{\mbox {h.c.}}\Big{]}\, , \label{HDOP-WIL}\\
\hspace{-.5cm}&& O_{6}^{R\, i} = \frac{1}{2} \rho
\Big{[}\bar Q_R\overleftarrow {\cal D}_\mu\frac{\tau^i}{2}\Phi^\dagger {\cal D}_\mu Q_L
-{\mbox {h.c.}}\Big{]}\, .
\label{HDOP-WIR}
\eeqn
Following the standard analysis of refs.~\cite{Bochicchio:1985xa,Testa:1998ez} and given the symmetries of ${\cal L}_{\rm toy}$ (see sect.~(\ref{sec:GSM})), one concludes that the operators~(\ref{HDOP-WIL}) and~(\ref{HDOP-WIR}) mix with two $d = 4$ operators, plus a set of six-dimensional ones that we will globally denote by $[O_{6}^{L\, i}]_{sub}$ and $[O_{6}^{R\, i}]_{sub}$, respectively~\footnote{We do not need to resolve the mixing among the different $d=6$ operators, as they only yield negligible O($b^2$) effects.}, according to the formulae
\beqn
\hspace{-1.2cm}&& O_{6}^{L\, i} = 
\Big{[} O_{6}^{L\, i}  \Big{]}_{sub} +
\frac{Z_{\tilde{J}}-1}{b^{2}}\partial_\mu\tilde J^{L\, i}_\mu
-\frac{\bar\eta}{b^{2}}\Big{[}\bar Q_L\frac{\tau^i}{2}\Phi Q_R
-{\mbox {h.c.}}\Big{]} + \ldots
\label{O6L-MIX} \\
\hspace{-1.2cm}&& O_{6}^{R\, i}  = 
\Big{[} O_{6}^{R\, i}  \Big{]}_{sub} +
\frac{Z_{\tilde{J}}-1}{b^{2}}\partial_\mu\tilde J^{R\, i}_\mu
-\frac{\bar\eta}{b^{2}}\Big{[}\bar Q_R\frac{\tau^i}{2}\Phi^\dagger Q_L
-{\mbox {h.c.}}\Big{]} + \ldots \, ,
\label{O6R-MIX}
\eeqn
where $Z_{\tilde{J}}$ and $\bar\eta$ are functions of the bare parameters entering~(\ref{SULL}). Details on the symmetry arguments leading to eqs.~(\ref{O6L-MIX}) and~(\ref{O6R-MIX}) are given in Appendix~B. Here we just note that in deriving these equations the conservation laws $\partial_\mu J^{L\, i}_\mu =0$ and $\partial_\mu J^{R\, i}_\mu =0$ have been used to eliminate from the mixing pattern the purely $\Phi$-dependent operators
$ \partial_\mu\mbox{\rm tr}\big{[}\Phi^\dagger\frac{\tau^i}{2}\partial_\mu\Phi - (\partial_\mu\Phi^\dagger)\frac{\tau^i}{2}\Phi\big{]} $   and
$\partial_\mu \mbox{\rm tr}\big{[}(\partial_\mu\Phi^\dagger)\Phi\frac{\tau^i}{2} - \frac{\tau^i}{2}\Phi^\dagger \partial_\mu\Phi\big{]} $.
Ellipses in the r.h.s.\ of eqs.~(\ref{O6L-MIX}) and~(\ref{O6R-MIX}) denote possible NP contributions to operator mixing. They are the main focus of this paper and the circumstances of their possible occurrence will be discussed in next section.
  
\section{$\tilde \chi_L \times \tilde \chi_R$ symmetry enhancement and naturally light fermion mass}
\label{sec:FINE2}

The physics of the model~(\ref{SULL}) with enhanced $\tilde \chi_L \times \tilde \chi_R$ symmetry (see eq.~(\ref{ETACR-RHO})) is drastically different depending on whether the parameter $\mu^2_0$ is such that ${\cal V}(\Phi)$ has a unique minimum (Wigner phase of the $\chi_L \times \chi_R$ symmetry) or whether ${\cal V}(\Phi)$ develops the typical ``mexican hat'' shape (Nambu--Goldstone phase). In the next subsections we discuss in detail the physical consequences of these two possible scenarios, and we shall argue that in the second case indeed a NP contribution arises in the r.h.s.\ of eqs.~(\ref{O6L-MIX}) and~(\ref{O6R-MIX}).

\subsection{The Wigner phase of $\chi_L \times \chi_R$ symmetry and the $\eta_{cr}$ definition}
\label{sec:WP}

If $\mu_0^2$ is such that ${\cal V}(\Phi)$ has a single minimum, one gets $\langle \Phi \rangle = 0$. In this situation we expect the $\Phi$-field to effectively provide no seed for dynamical $\tilde\chi_L \times \tilde\chi_R$-symmetry breaking (D$\tilde\chi$SB). As a consequence no NP terms (i.e.\ ellipses) of the type discussed in sect.~(\ref{sec:NGP}) are expected to occur in the mixing pattern of eqs.~(\ref{O6L-MIX}) and~(\ref{O6R-MIX}), that is thus assumed to be just the one visible in perturbation theory. Indeed, we shall see below that NP effects associated to D$\tilde\chi$SB necessarily involve a non-analytic function of the $\Phi$-field that is not well defined if $\langle \Phi \rangle = 0$.

\begin{figure}[htbp]    
\centerline{\includegraphics[scale=0.50,angle=0]{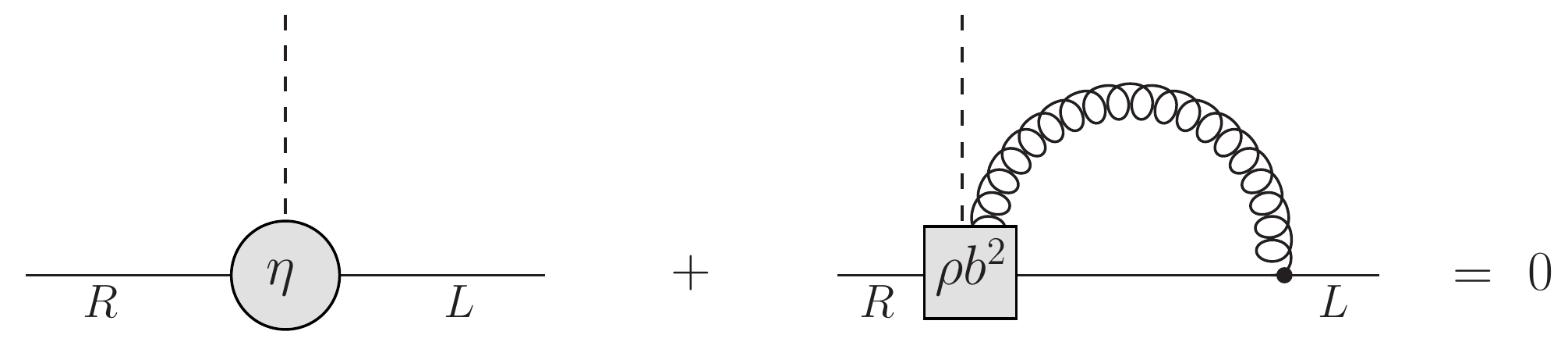}} 
\caption{\small{The Yukawa term cancellation mechanism determining $\eta_{cr}$ in the Wigner phase.}}
\label{fig:CANC}    
\end{figure}

The critical value of $\eta$ at which (up to irrelevant O($b^2$) terms) the transformations $\tilde \chi_L \times \tilde \chi_R$ become a symmetry of the theory can be consistently determined by {\em imposing} the validity of the renormalized WTIs 
\beqn
\hspace{-1.4cm}&&\partial_\mu \langle Z_{\tilde J}\tilde J^{L\, i}_\mu(x) \,\hat O(0)\rangle\Big{|}_{cr} 
= \langle \tilde\Delta_{L}^i\hat O(0)\rangle\Big{|}_{cr}\delta(x)+{\mbox O}(b^2)\, ,\label{CTLTI-RCR}\\
\hspace{-1.4cm}&&\partial_\mu \langle Z_{\tilde J}\tilde J^{R\, i}_\mu(x) \,\hat O(0)\rangle\Big{|}_{cr} = \langle \tilde\Delta_{R}^i \hat O(0)\rangle\Big{|}_{cr}\delta(x)
+{\mbox O}(b^2)\label{CTRTI-RCR}\, .
\eeqn
Inserting eqs.~(\ref{O6L-MIX}) and~(\ref{O6R-MIX}) - with ellipses now set to zero - in the WTIs~(\ref{CTLTI}) and~(\ref{CTRTI}), we see that taking $\eta$ equal to the solution of the equation 
\beq
\eta= \bar\eta(g_s^2,\rho,\lambda_0,\eta) \,\Longrightarrow \,\eta=\eta_{cr}(g_s^2,\rho,\lambda_0) 
\label{ETACR-RHO}
\eeq 
makes the $\tilde\chi_{L/R}$-variation of the $d=4$ Yukawa term to cancel the $d=4$ operator that mixes with $-b^2 O_6^{L/R\, i}$ (the latter we recall is the $\tilde\chi_{L/R}$-variation of the Wilson-like term in the action). As a consequence, in the r.h.s.\ of the WTIs~(\ref{CTLTI}) and~(\ref{CTRTI}) only genuinely $d \geq 6$ {\it subtracted} operators are left, which contribute irrelevant O($b^2$) cutoff artefacts. In fig.~\ref{fig:CANC} we schematically illustrate the Yukawa term cancellation mechanism that determines the value of $\eta_{cr}$ in the Wigner phase. 

One can check that $\eta_{cr}$ is odd under a change of sign of $\rho$, as it follows from the invariance of ${\cal L}_{\rm toy}$ under $\tilde{\cal R}_5 \times (\rho \to -\rho) \times (\eta \to -\eta)$, where $\tilde{\cal R}_5$ %(see footnote\,$^8$) 
is a $Z_2$-subgroup of $\tilde{\chi}_L \times \tilde{\chi}_R$, corresponding to the non-anomalous discrete transformation
\beq
 Q \to Q'= \gamma_5 Q \, \qquad \bar Q \to \bar Q' = -\bar Q\gamma_5\, .
\eeq
The fact that the same value of $\eta_{cr}$ makes both eqs.~(\ref{CTLTI-RCR}) and~(\ref{CTRTI-RCR}) to hold is a consequence of the invariance of ${\cal L}_{\rm toy}$ under parity, $P$. Furthermore we note that $\eta_{cr}=\eta_{cr}(g_s^2,\rho,\lambda_0)$ does not depend on the scalar field (squared) mass~\footnote{The squared mass of $\Phi$ undergoes both additive and multiplicative renormalization. The parameter $\mu_0^2$ is related to its renormalized counterpart, $\hat\mu_\Phi^2$, by $\hat\mu_\Phi^2 = Z_{\Phi^\dagger\Phi}^{-1} [\mu_0^2 - \tau b^{-2}]$, with $\tau$ a dimensionless function of $g_s^2$, $\lambda_0$, $\eta$ and $\rho$. Since $\eta_{cr}$ can only be a function of dimensionless bare parameters, it can depend on the scalar squared mass only via the quantity $b^2 Z_{\Phi^\dagger\Phi} \hat\mu_\Phi^2 = b^2 \mu_0^2 - \tau$, i.e.\ a negligible O($b^2$) effect.}.

In conclusion, if one sets $\eta=\eta_{cr}$ in~(\ref{SULL}), the transformations $\tilde{\chi}_L \times\tilde{\chi}_R$ are promoted to symmetries of the action up to irrelevant O($b^2$) cutoff effects. Recalling the form of the exact symmetries $\chi_L$ (eq.~\ref{GTWT}) and $\chi_R$ (eq.~\ref{GTCT}), this implies that also the transformations
\beq
\chi_L^{\Phi} : \qquad \Phi\to\Omega_L\Phi  
\, , \qquad \Omega_L\in {\mbox{SU}}(2)_L   \label{CHiPHI-L}
\eeq
and 
\beq
\chi_R^{\Phi} : \qquad \Phi\to\Phi\Omega_R^\dagger  
\, , \qquad
\Omega_R\in {\mbox{SU}}(2)_R \label{CHiPHI-R}
\eeq
become symmetries (up to O($b^2$) effects). To this order the corresponding currents $J_\mu^{L\,i} - \tilde J_\mu^{L\,i}$ and $J_\mu^{R\,i} - \tilde J_\mu^{R\,i}$, that involve 
only scalar fields (see eqs.~(\ref{CCL})-(\ref{CCR}) and~(\ref{JCLT})-(\ref{JCRT})) are conserved. As a result, when the condition~(\ref{ETACR-RHO}) is fulfilled, the scalar field gets actually decoupled (up to O($b^2$) artefacts) from fermion and gauge boson degrees of freedom. More precisely, at the critical value of $\eta$, the newly enforced $\tilde \chi_L\times\tilde\chi_R$ invariance implies (up to O($b^2$)) the same set of relations among correlators involving only fermions and gluons as the exact $\chi_{L} \times \chi_{R}$ symmetry. 

The local part of the 1PI effective Lagrangian of the theory in the Wigner phase ($\hat \mu_\Phi^2>0$) takes the form
\beqn
\hspace{-1.6cm}&& L_4^{Wig} =  \frac{1}{4}(F\cdot F)+\bar Q_L\Dslash Q_L+\bar Q_R\Dslash \,Q_R + \nonumber \\
\hspace{-1.6cm}&& +\frac{1}{2}{\tr}\big{[}\partial_\mu\Phi^\dagger\partial_\mu\Phi\big{]} +
\frac{\hat\mu_\phi^2}{2}{\tr}\big{[}\Phi^\dagger\Phi\big{]}+\frac{\hat\lambda}{4}\big{(}{\tr}\big{[}\Phi^\dagger\Phi\big{]}\big{)}^2 \, .
\label{L4Wig}
\eeqn
The expression of $L_4^{Wig}$ is completely determined by symmetry requirements and for this reason is sometimes called the ``target theory'', i.e.\ the theory one is aiming at. In the case at hand, besides the obvious gauge, Lorentz and $C$, $P$, $T$ symmetries, its form is constrained by requiring invariance under $\chi_{L} \times \chi_{R}$ transformations as well as $\tilde \chi_L\times\tilde\chi_R$. The expression~(\ref{L4Wig}) clearly shows that scalars are completely decoupled from fermions and gluons. 
From a different vantage we can also say that, once the details of the UV-regularization have been fully specified, the correlators of the UV-regulated model, computed with the Lagrangian ${\cal L}_{\rm toy}$ admit a small-$b$ Symanzik expansion in terms of correlators of the formal model defined by the $d=4$ Lagrangian~(\ref{L4Wig}).

\subsection{The Nambu--Goldstone phase of the $\chi_L \times \chi_R$ symmetry and the effects of D$\tilde\chi$SB}
\label{sec:NGP}

We now want to investigate the physical properties of the model that is obtained if the parameter $\mu_0^2$ in ${\cal L}_{\rm toy}$ is brought to a value such that ${\cal V}(\Phi)$ develops a double-well shape, while the dimensionless Yukawa coupling is kept at the critical value, $\eta_{cr}$, that was determined (at a value of $\mu_0^2$ at which the model is) in the Wigner phase. Since, as we remarked above, $\eta_{cr}$ is independent from the renormalized scalar mass $\hat\mu_\Phi^2$, its value is not affected by a change of sign of $\hat\mu_\Phi^2$ (i.e.\ if one now takes $\mu_0^2 - \tau b^{-2}<0$).

With the $\chi_L \times \chi_R$ symmetry realized {\it \`a la} Nambu--Goldstone the physics of the model is much more interesting than the situation we have discussed in the previous section. To see what happens we expand, as usual, the scalar field around its vev by writing 
\begin{equation}
\Phi(x) = (v+ \sigma(x)) 1_{2\times 2} + i \vec{\pi}(x) \vec{\tau} \, , 
\label{PHIAROUNDV}
\end{equation}
where $\vec{\pi}$ is a triplet of massless pseudoscalar Nambu--Goldstone bosons and $\sigma$ is a scalar of mass $m_\sigma = {\rm O}(v)$. It is worth recalling that in the Nambu--Goldstone vacuum defined by the expansion~(\ref{PHIAROUNDV}) the $\chi_L\times\chi_R$-symmetry of ${\cal L}_{\rm toy}$ is reduced to its diagonal sub-group, $\chi_V$.

In the following we shall argue that a natural choice is to take $v$ much larger than the RGI scale of the theory, $v\gg \Lambda_{s}$, but still $\ll b^{-1}$. The compelling reason for the inequality $v\gg \Lambda_{s}$ will be spelled out in the point 5) of sect.~\ref{sec:THEOREMA}. 

We immediately note that, ignoring the fluctuations of $\Phi$ around its vev, the $d=6$ term ${\cal L}_{Wil}$ with $b^2 v \to ar$ looks very much like the $d=5$ Wilson term in LQCD. We may then expect that the residual $\tilde \chi_L\times\tilde\chi_R$-breaking terms left-over at $\eta_{cr}$, where ${\cal L}_{Wil}$ is (partially) compensated by ${\cal L}_{Yuk}$, will trigger the phenomenon of D$\tilde \chi$SB, just as it happens in LQCD with Wilson fermions, where chiral symmetry plays the same r\^ole as the $\tilde \chi_L\times\tilde\chi_R$-symmetry in the present model. Indeed in the familiar case of LQCD, owing to the residual explicit O($a$) breaking of chirality, we know that the phenomenon of spontaneous chiral symmetry breaking occurs even when $m_0$ is set at $m_{cr}$ (see eq.~(\ref{CRM1})) and the Wilson term gets (partially) compensated by the mass term~\footnote{The well-known fact that the (critical) Wilson term can trigger the phenomenon of spontaneous chiral symmetry breaking in LQCD is incorporated in the formalism of Wilson chiral perturbation theory~\cite{SingShar,RupSho02,BarRupSho03} and gives rise to the peculiar lattice scenarios of S$\chi$SB~\cite{SingShar,Aoki83,Muns04,SharWu}, differing from continuum QCD by O($a^2$) effects, that are actually observed in numerical simulations (see e.g.\ refs.~\cite{Farch04,Sharpe05}).}.

In order to determine the structure and the properties of the {\it critical} theory in the double-well situation, we need to analyse how NP terms coming from D$\tilde\chi$SB effects can affect correlators and in particular the building blocks that enter the quark self-energy diagrams. Thus among others, we will focus on the small-$b$ expansion of the gluon-gluon-scalar, $Q_{L/R}$-$\bar Q_{L/R}$-scalar, $Q_{L/R}$-$\bar Q_{L/R}$-gluon-scalar correlators that take the form (as we have observed before, terms odd in $b$ in the SLEL of the model are excluded by the ${\cal L}_{\rm toy}$ symmetries) 
\begin{eqnarray}
\hspace*{-.3cm}&&\langle O(x,x',...)\rangle\Big{|}^R=\langle O(x,x',...)\rangle\Big{|}^F+\nn\\
\hspace*{-.3cm}&& \quad-b^2\langle O(x,x',...) \int d^4 z \, [ \, L_6^{/\!\!\!\tilde\chi}
+ L_6^{\tilde\chi}\,](z) \rangle\Big{|}^F + {\rm O}(b^4)  \, , \label{ALCSYM} \\
\hspace*{-.3cm}&&O(x,x',...) \Leftrightarrow A_\mu^b(x) A_\nu^c (x')\sigma(y) \, ,\; Q_{L/R}(x) \bar Q_{L/R}(x') \sigma(y) \, ,\nn\\
\hspace*{-.8cm}&&\phantom{O(x,x',...) \Leftrightarrow}\,Q_{L/R}(x) \bar Q_{L/R}(x') \sigma(y) A_\mu^b(y') \, ,\ldots \label{ALCSYMO}
\end{eqnarray}
where $L_6^{/\!\!\!\tilde\chi}$ is the $d=6$ $\tilde\chi$-breaking SLEL operator and $L_6^{\tilde\chi}$ the $d=6$ $\tilde\chi$-conserving one. The label $|^R$ in eq.~(\ref{ALCSYM}) means that expectation values are taken in the UV-regulated ${\cal L}_{\rm toy}$ theory, while the label $|^F$ means that expectation values are taken in the ``formal" theory. The latter should be identified with the ``target theory" of the critical model in the Nambu--Goldstone phase. Its Lagrangian (which also coincides with the $d=4$ piece of the SLEL) can be represented by the formula
\beqn
\hspace{-1.6cm}&& L_4^{NG} =L_4^{Wig}\Big{|}_{\hat \mu_\Phi^2<0}  +\epsilon \Big{(} \bar Q_L\Phi Q_R +\bar Q_R\Phi^\dagger Q_L \Big{)}\Big{|}_{\epsilon\to 0+}\, ,
\label{L4NG}
\eeqn
where the last term is introduced to have  the phenomenon of D$\tilde \chi$SB formally 
implemented in the Nambu--Goldstone phase of the  theory.

One checks that gauge symmetry and Lorentz invariance together with dimensional arguments make  the expectation values of the operators~(\ref{ALCSYMO}) (first term in the r.h.s.\ of eq.~(\ref{ALCSYM})) to vanish in the formal theory. 

\subsubsection{Symanzik expansion}
\label{sec:SENGP}

The analysis of the Symanzik expansion that follows is analogous to the one presented in sect.~\ref{sec:SYMEXP}. Indeed, like in LQCD, the O($b^2$) terms with the insertion of $L_6^{/\!\!\!\tilde\chi}$ would vanish were it not for the NP phenomenon of D$\tilde\chi$SB. 
The resulting NP contributions to the gluon-gluon-scalar, $Q_{L/R}$-$\bar Q_{L/R}$-scalar, $Q_{L/R}$-$\bar Q_{L/R}$-gluon-scalar vertices will have the form 
\begin{eqnarray}
\hspace{-0.6cm}&&\Delta \Gamma_{AA\Phi}^{bc\,\mu\nu}(k,\ell) \Big{|}^R \!\! =  b^2\Lambda_{s} \alpha_s\,\frac{\delta^{bc}}{2} T_{\mu\nu}
\,F_{AA\Phi}\Big{(}\frac{\Lambda_s^2}{{\rm mom}^2} \Big{)}
\, ,\label{DELTAAAF}\\
\hspace{-0.6cm}&&\Delta \Gamma_{Q \bar Q \Phi}(k,\ell) \Big{|}^R \!\! =  b^2\Lambda_{s} \,
\alpha_s\, \frac{i}{2} \gamma_\mu (2k+\ell)_\mu 
 \,F_{Q \bar Q \Phi}\Big{(}\frac{\Lambda_s^2}{{\rm mom}^2} \Big{)} 
\, ,\label{DELTAQQF} \\
\hspace{-0.6cm}&&\Delta \Gamma_{Q \bar Q A \Phi}^{b,\mu}(k,\ell,\ell') \Big{|}^R \!\! =  b^2\Lambda_{s} \,
\alpha_s\, i g_s \lambda^b \gamma_\mu 
 \,F_{Q \bar Q A \Phi}\Big{(}\frac{\Lambda_s^2}{{\rm mom}^2} \Big{)}\, ,\label{DELTAQQAF} 
\end{eqnarray}
respectively, where we have set 
\beq
T_{\mu\nu}=[k(k+\ell)\delta_{\mu\nu}-k_\mu (k+\ell)_\nu]+[\mu\to\nu]\, ,\label{TMN}
\eeq 
and ``mom" stands for any one of the kinematically appropriate momenta in the set $\{k,\ell,\ell', ... ,\ell'+\ell,k+\ell\}$. As in LQCD, the factor $\alpha_s$ comes from the fact that the quark or gluon line emitted from the $L_6^{/\!\!\!\tilde\chi}$ vertex has to be absorbed somewhere in the diagram. We also note that an analysis of the structure of the Schwinger--Dyson equations shows that the NP corrections to the vertices under consideration start to appear precisely at first order in the gauge coupling $\alpha_s$. As before, at this stage we leave unspecified the scale at which the gauge running coupling should be evaluated. 

\begin{figure}[htbp]
	\centering  
\includegraphics[width=0.6\linewidth]{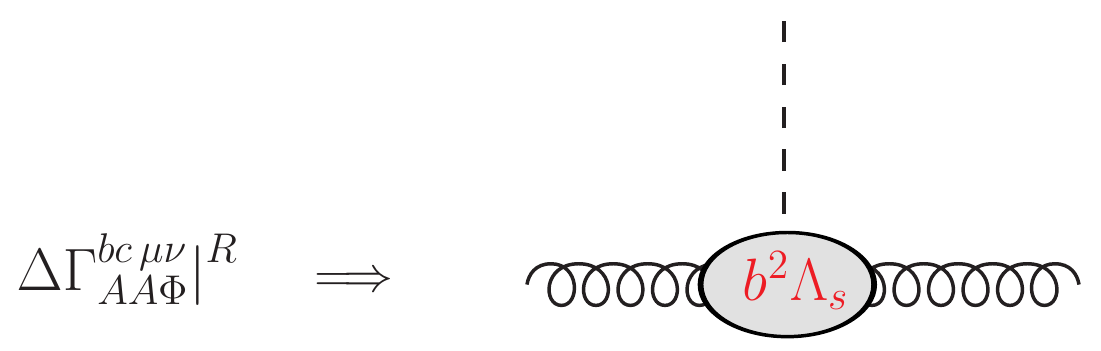}
\includegraphics[width=0.6\linewidth]{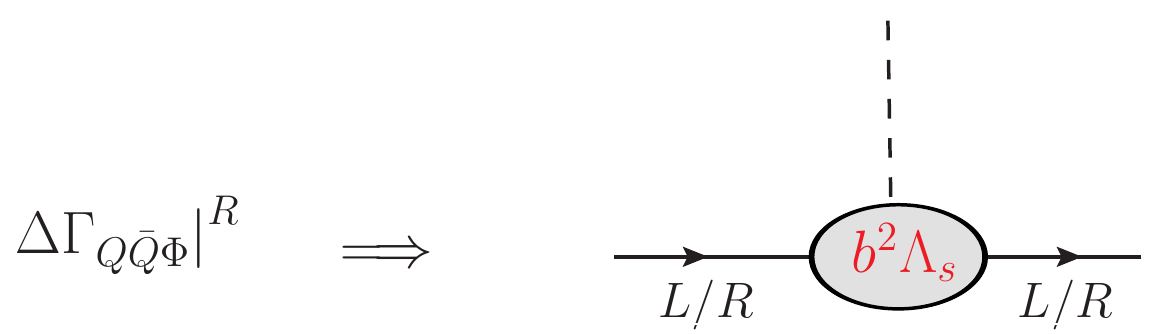}    
\includegraphics[width=0.65\linewidth]{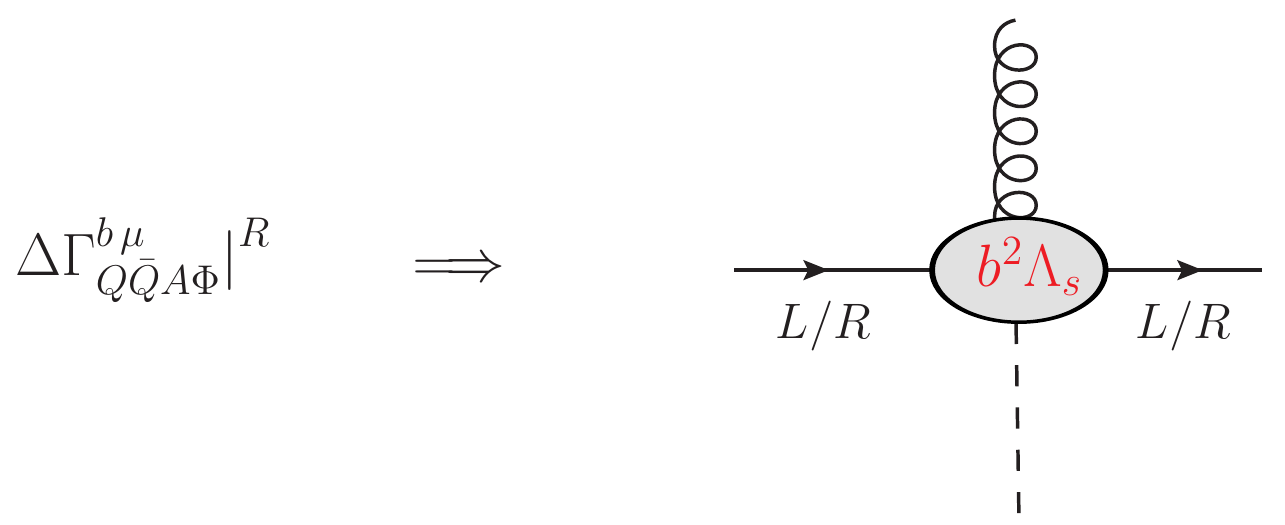}	
\caption{\small{The NP terms of order $b^2\Lambda_s \alpha_s$ contributing to the left, central and right panel of fig.~\ref{fig:LEAD-NPMIX-DIAG}, respectively.}}
\label{fig:FIG671}
\end{figure}

The scalar form factors $F_{AA\Phi}$, $F_{Q \bar Q \Phi}$ and $F_{Q \bar Q A \Phi}$ are dimensionless functions with a non-trivial dependence on the $\Lambda_{s}^2/{\mbox{mom}}^2$ ratios. In the following we shall represent the above O($b^2$) NP contributions  to gluon-gluon-scalar, $Q_{L/R}$-$\bar Q_{L/R}$-scalar, $Q_{L/R}$-$\bar Q_{L/R}$-gluon-scalar vertices with the symbols depicted in the right panels of fig.~\ref{fig:FIG671}.

Standard arguments \`a la Symanzik imply that small-$b$ expansions like those in~(\ref{ALCSYM}) are expected to be valid for squared momenta much smaller than the UV-cutoff, $b^{-2}$. Like in LQCD, we assume that the NP effects encoded in eqs.~(\ref{ALCSYM}) to~(\ref{DELTAQQAF}) persist up to mom$^2 ={\mbox{O}}(b^{-2})$, and {\it conjecture} the asymptotic behaviour 
\begin{eqnarray}
&&F_{AA\Phi}\Big{(} \frac{\Lambda_{s}^2}{{\rm mom}^2} \Big{)} 
  \stackrel{{\rm mom}^2\to\infty}\longrightarrow  H_{AA}
\, , \label{FAA}\\
&&F_{Q \bar Q \Phi}\Big{(} \frac{\Lambda_{s}^2}{{\rm mom}^2} \Big{)} 
\stackrel{{\rm mom}^2\to\infty}\longrightarrow H_{Q \bar Q} \, ,\label{FQQ}\\
&&F_{Q \bar Q A \Phi}\Big{(} \frac{\Lambda_{s}^2}{ {\rm mom}^2 } \Big{)}
\stackrel{{\rm mom}^2\to\infty}\longrightarrow H_{Q \bar Q} \, ,\label{FAAF}
\end{eqnarray}
where $H_{AA}$ and $H_{Q \bar Q}$ are O(1) constants and the last two limits are related by gauge invariance. 

\subsection{Dynamical quark mass generation and $\tilde\chi_L \times \tilde\chi_R$ WTIs}
\label{sec:DLMM}

With the building blocks provided by the NP O($b^2$) corrections to the gluon-gluon-scalar, $Q_{L/R}$-$\bar Q_{L/R}$-scalar and $Q_{L/R}$-$\bar Q_{L/R}$-scalar-gluon vertices given in eqs.~(\ref{DELTAAAF}), (\ref{DELTAQQF}) and~(\ref{DELTAQQAF}), we are in position to compute the leading fermion self-energy ``diagrams'' and the structure of the NP mixing pattern of $O_{6}^{L\, i}$ and $O_{6}^{R\, i}$ (see eqs.~(\ref{O6L-MIX}) and~(\ref{O6R-MIX}))~\footnote{Actually, like in LQCD, NP corrections appear also in other $n$-point correlators. The $\Delta{\cal L}|_{ad\,hoc}$ that would describe all these effects is much more complicated than the formula we gave for LQCD. Here we only report for illustration the part of $\Delta{\cal L}|_{ad\,hoc}$ that is relevant for the calculation of the ``diagrams'' displayed in fig.~\ref{fig:LEAD-NPMIX-DIAG}.

Based on the (non-spontaneously broken) symmetries of the model~(\ref{SULL}), 
to leading order in $g_s^2$ (and $b^2$) the terms necessary to describe the NP O($b^2$) terms in the Symanzik expansion of the correlators~(\ref{ALCSYMO}) can be compactly encoded in the expression  
\begin{eqnarray}
\hspace{-0.7cm}&&\Delta{\cal L}\Big{|}_{ad\,hoc}=\frac{b^2}{2}\Lambda_{s}\alpha_s 
{\rm tr}[\Phi+ \Phi^\dagger] \Big{[}\frac{H_{AA}}{4} (F\cdot F)
+ H_{Q \bar Q} (\bar Q {\slash{\cal D}} Q) \Big{]}  + \ldots \, .\nn
\end{eqnarray}
We note again that the augmented Lagrangian ${\cal L}_{\rm toy} + \Delta{\cal L}|_{ad\,hoc}$ should be only seen as a useful tool to get insights about the structure of NP contributions in correlators, as it reproduces the NP O($b^2$) vertex contributions we inferred from the Symanzik expansion and provides a way of embedding them in a sort of ``non-perturbatively augmented Feynman rules''. Naturally, complete and reliable computations can only be performed by means of numerical simulations of the fundamental theory represented by the Lagrangian~(\ref{SULL}).}.
This way of estimating NP effects in certain vertices of the model~(\ref{SULL}) can
be justified looking at the structure of the relevant Schwinger--Dyson equations.

\subsubsection{Dynamical NP mass generation}
\label{sec:NPMG}

\begin{figure}[htbp]   
\vspace{-.5cm}
\hspace{-1.cm} 
\includegraphics[scale=0.75,angle=-0]{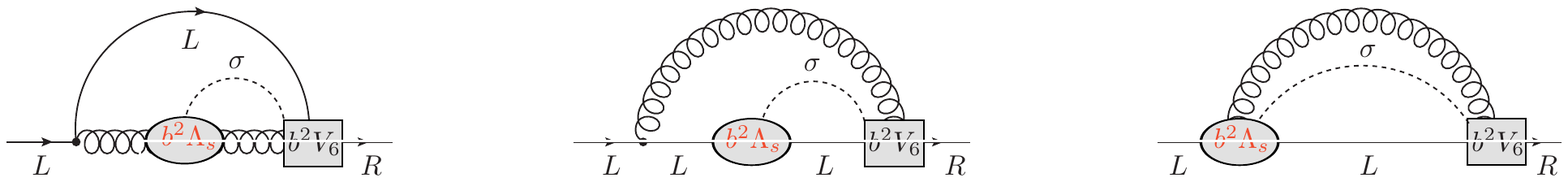} 
\vspace{-20.5cm}
\caption{\small{Typical lowest order ``diagrams'' giving rise to dynamically generated quark mass terms ($L$ and $R$ are fermion helicity labels). The grey blob represents the NP  $b^2\Lambda_s\alpha_s$ effect embodied in eqs.~(\ref{DELTAAAF}), (\ref{DELTAQQF}) and~(\ref{DELTAQQAF}), respectively. The grey box represents the insertion of the Wilson-like vertex stemming from ${\cal L}_{Wil}$. The dotted line represents the propagation of a $\sigma$ particle.}}
\label{fig:LEAD-NPMIX-DIAG}
\end{figure}

For illustration in fig.~\ref{fig:LEAD-NPMIX-DIAG} we report a few self-energy ``diagrams'' that give rise to finite O($g^2_s\alpha_s\Lambda_{s}$) NP contributions to the fermion mass.

The finiteness of these contributions is apparent from a straightforward counting of loop momenta in the graphs. For instance, with reference to the central panel of fig.~\ref{fig:LEAD-NPMIX-DIAG} and neglecting external (compared to loop) momenta, one finds a double integral with factors $1/k^2$ and $1/(\ell^2 + m_{\sigma}^2)$ from the standard gluon and $\sigma$ propagators, the factors $\gamma_\mu k_\mu/k^2$ and $\gamma_\nu (k+\ell)_\nu/(k+\ell)^2$ for the quark propagators, a factor $b^2(k+\ell)_\lambda$ from the ${\cal L}_{Wil}$ derivative coupling and a factor $b^2 \alpha_s \Lambda_s(2k+\ell)_\rho \gamma_\rho$ from the NP vertex $\Delta \Gamma_{Q \bar Q \Phi}(k,\ell)|^R$. Putting everything together, one gets in the $b\to 0$ limit (similarly to eq.~(\ref{CONS})) a finite fermion mass term of the order 
\beqn
\hspace{-.8cm}&&b^4g_s^2\alpha_s\Lambda_s\!\!\int^{1/b}\!\!d^4k\int^{1/b}\!\!d^4\ell\,
\frac{1}{k^2} \gamma_\lambda \frac{\gamma_\mu k_\mu}{k^2} 
\frac{(2k+\ell)_\rho \gamma_\rho }{\ell^2+m^2_{\sigma}}
 \frac{\gamma_\nu (k+\ell)_\nu}{(k+\ell)^2} (k+\ell)_\lambda \sim 
\nn\\
\hspace{-.8cm}&&\quad\sim g_s^2\alpha_s\Lambda_s\, ,
\label{MASSB}
\eeqn 
as the overall $b^4$ multiplicative factor is compensated by the quartic divergency of the two-loop integrals. The diagrams in fig.~\ref{fig:LEAD-NPMIX-DIAG} represent a subset of all the lowest order terms contributing to the fermion self-energy, namely those where only one $\sigma$ propagator appears. To the same lowest order in $g_s^2$ there are infinitely many other contributions coming from diagrams that take into account the self-interaction of the $\Phi$ field and include in general scalar ($\sigma$ and/or $\pi$) loops.

Unlike the case of LQCD, we are not going to present the alternative argument for NP fermion self-mass generation that relies on the use of the spectral density of the average fermion Dirac operator (in the vacuum~(\ref{PHIAROUNDV}) of the Nambu--Goldstone phase). In fact, in the UV-regulated ${\cal L}_{\rm toy}$ model we should deal with an at least three-loop calculation (see figs~\ref{fig:FIG671} and~\ref{fig:LEAD-NPMIX-DIAG}) and such an effort appears to be beyond the scope of this speculative paper.

Nevertheless to be able to interpret the finite term we have just identified as a {\it bona fide} quark mass we ought to prove the following statements.  

	1) No extra O($v$) quark mass is left over as a consequence of the Higgs mechanism because a term of that kind would completely obscure the NP contribution~(\ref{MASSB}) in case $v\gg\Lambda_{s}$, or make it of little interest for predicting the value of the quark mass, in case $v \sim \Lambda_{s}$.
	
	2) A $\chi_L\times\chi_R$-invariant NP mass term of the magnitude~(\ref{MASSB}) must exist that is endowed with the correct symmetry properties to appear in the effective Lagrangian of the model in its Nambu--Goldstone phase.
	
	3) The NP fermion mass term is renormalization scale independent and its chiral variation can be accommodated in the r.h.s.\ of the restored $\tilde\chi_L\times\tilde\chi_R$ WTIs. 

\begin{figure}[htbp]    
\centerline{\includegraphics[scale=0.50,angle=0]{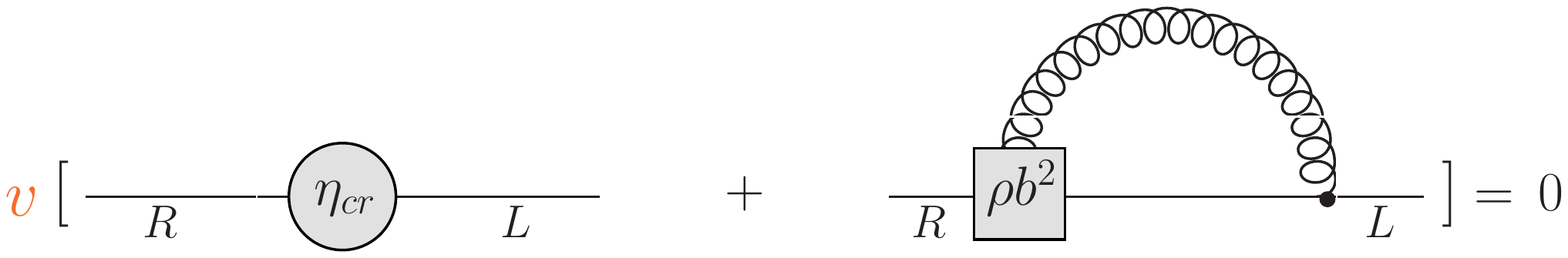}}   
\vspace{-12cm}  
\caption{\small{The mechanism for the O($v$) quark mass cancellation in the Nambu--Goldstone phase.}}
\label{fig:VCANC}  
\end{figure}

We discuss the first of these three issues in this subsection and leave the other two for the next two subsections. The first statement is proved by observing that in the Nambu--Goldstone phase the equation determining $\eta_{cr}$ becomes just a condition for the cancellation of the $v (\bar Q_{R}Q_L+\bar Q_{L} Q_R)$ quark mass term (compare fig.~\ref{fig:VCANC} with fig.~\ref{fig:CANC}).

\subsubsection{The mass term}
\label{sec:INU}

Proving the other two statements looks much more challenging and interesting because 
a naive mass term of the kind $\propto\Lambda_s(\bar Q_{R}Q_L+\bar Q_{L} Q_R)$ in 
the effective Lagrangian is forbidden by the exact $\chi_L\times\chi_R$-invariance of ${\cal L}_{\rm toy}$. 

The solution of this seemingly insoluble problem is of a NP nature and requires introducing the field 
\beq 
U = \frac{\Phi}{\sqrt{\Phi^\dagger \Phi}}\, .\label{U}
\eeq 
$U$ is a dimensionless non-analytic function of $\Phi$ that has the same transformation properties as the latter under $\chi_L \times \chi_R$ and is well defined only if $\langle\Phi\rangle=v\neq 0$~\footnote{It may be worth noting that $U$ is the phase of $\Phi$ and can always be written in the form $U={\mbox{sign}}\,(v+\sigma)\exp{(i\vec \tau \vec \zeta/v)}\, ,\,\vec\zeta=\vec\pi\,[1+{\mbox{O}}(\sigma/v, \vec\tau \vec\pi/v)]$.}. In terms of $U$ one can construct the desired NP $\chi_L \times \chi_R$-invariant quark mass-like term which reads~\footnote{Actually one cannot exclude that eq.~(\ref{NPMASS}) has the more general form $C_1 \Lambda_s [\bar Q_L U Q_R + \bar Q_R U^\dagger Q_L]{\cal P}$, where the factor ${\cal P} = {\cal P}(v^{-2} \Phi^\dagger \Phi)$ is a $\chi_L \times \chi_R$-invariant function of $v^{-2} \Phi^\dagger \Phi$ such that ${\cal P}|_{\Phi^\dagger \Phi = v^2 1\!\!1} = 1\!\!1$. Like $U$, ${\cal P}$ is well defined only if $v> 0$ (i.e.\ for $\hat\mu_\Phi^2 <0$). We stress that the appearance  of $U$ and possibly ${\cal P}$ in our formulae is necessary for describing the many other NP contributions, besides the ones shown in fig.~\ref{fig:LEAD-NPMIX-DIAG}, that arise because of the scalar field self-interaction. } 
\begin{equation}
C_1 \Lambda_s [\bar Q_L U Q_R + \bar Q_R U^\dagger Q_L]\, ,\label{NPMASS}
\end{equation}
where, in view of the result~(\ref{MASSB}), to leading order (LO) in $g_s^2$ one has
\begin{equation}
C_1\Big{|}_{\rm LO} = k_{\rm LO} g_s^2\, \alpha_s \, , \qquad
 k_{\rm LO} = {\rm O}(1) \, . \label{C1}
\end{equation}
The conjectured NP contributions to the quark self-energy imply the occurrence of the additional term $C_1 \Lambda_s [\bar Q_L U Q_R + \bar Q_R U^\dagger Q_L]$ in the {\em effective action} density of the model in its Nambu--Goldstone phase. The local piece should thus take the form (for $L_4^{Wig}$ see eq.~(\ref{L4Wig}) except that now $\hat \mu_\Phi^2 < 0$)
\beq
\Gamma_{\rm loc}^{NG} = L_4^{Wig}\Big{|}_{\hat \mu_\Phi^2 < 0}  + C_2\Lambda_s^2 \tr[\partial_\mu U^\dagger\partial_\mu U] + C_1 \Lambda_s [\bar Q_L U Q_R + \bar Q_R U^\dagger Q_L]  \, ,
\label{GAMMA4NG} \eeq
where, besides the ``mass term'' proportional to the RGI scale, we have also introduced a ``kinetic term'' for the non-linear field $U$ that cannot be excluded on the basis of symmetry considerations. Actually ``mixed kinetic terms" of the kind $\Lambda_s[\partial_\mu \Phi\partial_\mu U^\dagger +{\mbox{h.c.}}]$ are a priori possible in~(\ref{GAMMA4NG}). For generic values of $\rho$ and $v\gg \Lambda_s$ all the kinetic terms containing $U$ are negligibly small  corrections to the {\it bona fide} kinetic term of the scalar fields already present in $L_4^{Wig}$. In fact, in the limit $\Lambda_s/v\to 0$ all such kinetic term contributions of
NP origin as well as the $\tilde\chi_L \times \tilde\chi_R$-breaking terms in eq.~(\ref{GAMMA4NG}) that stem from the expansion of $U$ in terms of $\vec\pi$ and $\sigma $ fields (with the exception of the $\sim \Lambda_s \bar{Q} Q$ mass term) do disappear. 

As we shall see in ref.~\cite{FRNEW}, however, the situation turns out to be very different if electro-weak interactions are present. In this case implementing the $\tilde\chi_R\times \tilde\chi_L$ symmetry requires the tuning of also the parameter $\rho$. The critical value of $\rho$ is one where the standard, $\partial_\mu \Phi^\dagger \partial_\mu \Phi$, kinetic term  and the ``mixed" one, $\Lambda_s[\partial_\mu \Phi\partial_\mu U^\dagger +{\mbox{h.c.}}]$, are absent in $\Gamma_{\rm loc}^{NG}$. In these circumstances the kinetic term of the non-linear field $U$ cannot be neglected anymore and, indeed consistently, the $v$-dependence of the last two terms in eq.~(\ref{GAMMA4NG}) disappears. The reason is that, at the critical value of $\rho$ and $\eta$, in order to have the kinetic term of the $\pi$ fields canonically normalized, one is forced to rewrite everything in terms of the rescaled fields $\pi' \sim (\Lambda_s/v)\,\pi$.

Ending this section it is important to remark that the appearance in the game of the non-analytic field $U$ should not come too much as a surprise if one recalls that in QCD NP effects like the ones that make the chiral condensate non-vanishing are proportional to the sign of $m_q$, i.e.\ the sign of the coefficient of the chiral breaking term in the action. In LQCD at $m_0=m_{cr}$, the seed for NP S$\chi$SB effects is instead provided by the (critical) Wilson term. As a result such NP effects will be proportional to the sign of the Wilson coefficient, $r$. From this point of view it is illuminating to regard the Lagrangian~(\ref{SULL}) as a consistent model where the Wilson coefficient is elevated to a dynamical field, $\Phi$. Indeed, as we have shown above, the dynamically generated NP quark mass~(\ref{NPMASS}) turns out to be  proportional to $U=\exp[i {\rm Arg}(\Phi)]$ (times a factor ${\rm O}(\rho^2) \,{\rm sign}(\rho) $). 

\subsubsection{$\tilde\chi_L\times\tilde\chi_R$ WTIs, NP operator mixing and mass renormalization}
\label{sec:NPMIX}

The emergence of a NP mass term in the  $\tilde\chi_L \times \tilde\chi_R$ WTIs can be seen to be a consequence of the quadratically divergent mixing of the $d=6$ operators $O^{L\,i}_6$ and $O^{R\,i}_6$ with the non-perturbatively generated operators  
\beqn
C_1 \Lambda_s \Big(\bar Q_L \frac{\tau^i}{2} U Q_R - {\mbox{h.c.}} \Big) \, ,\qquad 
C_1 \Lambda_s \Big(\bar Q_R \frac{\tau^i}{2} U^\dagger Q_L-{\mbox{h.c.}}\Big)\, .
\label{NPcontrWTIR} 
\eeqn
This is precisely the possible NP mixing which was alluded to by the ellipses in
eqs.~(\ref{O6L-MIX}) and~(\ref{O6R-MIX}). Indeed, owing to $\chi_L \times \chi_R$ and other obvious symmetries, at $\eta=\eta_{cr}$ (see eq.~(\ref{ETACR-RHO})), the renormalized WTIs associated to the $\tilde\chi_L \times \tilde\chi_R$ transformations are conjectured to take the form~\footnote{To simplify formulae also in this section we systematically ignore the possible presence of the ${\cal P}(v^{-2}\Phi^\dagger\Phi)$ factor in the NP $\tilde\chi_L \times \tilde\chi_R$--breaking term.} 
\beqn
\hspace{-1.6cm}&&\partial_\mu \langle Z_{\tilde J}\tilde J^{L\, i}_\mu(x) \,\hat O(0)\rangle\Big{|}_{cr}=\langle \tilde\Delta_{L}^i\hat O(0)\rangle\Big{|}_{cr}\delta(x)+ \nn\\&&\qquad+C_1 \Lambda_s \langle (\bar Q_L \frac{\tau^i}{2} U Q_R - {\mbox{h.c.}})
\, \hat O(0)\rangle\Big{|}_{cr} +{\mbox O}(b^2)\, ,\label{CTLTI-RCR-NP}\\
\hspace{-1.6cm}&&\partial_\mu \langle Z_{\tilde J}\tilde J^{R\, i}_\mu(x) \,\hat O(0)\rangle\Big{|}_{cr}=\langle \tilde\Delta_{R}^i\hat O(0)\rangle\Big{|}_{cr}\delta(x)+\nn\\&&\qquad +C_1 \Lambda_s \langle (\bar Q_R \frac{\tau^i}{2} U^\dagger Q_L - {\mbox{h.c.}})
\, \hat O(0)\rangle\Big{|}_{cr} +{\mbox O}(b^2)\label{CTRTI-RCR-NP}\, .
\eeqn
Eqs.~(\ref{CTLTI-RCR-NP}) and~(\ref{CTRTI-RCR-NP}) show that in the critical theory, consistently with the form of the effective Lagrangian, $\Gamma_{\rm loc}^{NG}$ (eq.~(\ref{GAMMA4NG})), in the r.h.s.\ of these WTIs besides other NP contributions a quark mass term occurs that is proportional to $\Lambda_s$, and not to scalar field vev, $v=\langle\Phi\rangle$. To leading order in the gauge coupling we get (see eq.~(\ref{C1}))
\beq
m_Q^{dyn}\Big{|}_{\rm LO}= C_1\Big{|}_{\rm LO}\Lambda_s 
= k_{\rm LO} g^2_s\alpha_s \Lambda_s\, .
\label{mQ-dyn} \eeq

Since the $\tilde\chi$-currents $Z_{\tilde J}\tilde J^{L\, i}_\mu $ and $Z_{\tilde J}\tilde J^{R\, i}_\mu $ are UV-finite (as it follows e.g.\ from the fact that they are conserved up to O($b^2$) in the Wigner phase of the model), to be really entitled to interpret the coefficient $C_1 \Lambda_s $ in front of the last correlator in the r.h.s.\ of the WTIs~(\ref{CTLTI-RCR-NP}) and~(\ref{CTRTI-RCR-NP}) as a mass, we need to assume that this quantity is renormalized by the inverse of the renormalization constant of the 
operators~(\ref{NPcontrWTIR}). In Appendix~C we spell out necessary and sufficient conditions for this to happen. 

We note immediately that the assumed $\log b$--scaling properties of the coefficient $C_1$ are not in contradiction with the conclusions of ref.~\cite{Testa:1998ez} where it is proved that the power-divergent mixing coefficients are independent of the subtraction point. The reason is that in the case at hand NP effects provide a new scale $\Lambda_s$ (besides the subtraction point) which can give rise to the dependence on $\log b\Lambda_s$ of the coefficient $C_1$ that is indeed necessary to match the running with the UV-cutoff of the (matrix elements of the) operators~(\ref{NPcontrWTIR}). An interesting application of these considerations concerning the RG scaling properties of non-perturbatively generated fermion masses is discussed in sect.~\ref{sec:TECH} 

\subsubsection{Theoretical remarks}
\label{sec:THEOREMA}

A number of observations are in order here.

1) {\it The Goldstone boson issue} - The physics of the toy-model~(\ref{SULL}) in its Nambu--Goldstone phase is quite rich. In particular, we must notice that there are two sets of ``Goldstone bosons'', related to the two kinds of spontaneous symmetry breaking (S$\chi$SB) occurring in the model. The first set is associated to the spontaneous breaking of the exact $\chi_L\times\chi_R$-symmetry that is induced by a non-vanishing scalar vev. In a more realistic model, where $\chi_L$ is gauged to introduce electro-weak interactions, these Goldstone bosons will become the longitudinal electro-weak boson degrees of freedom. The second set of ``Goldstone bosons'' is associated to the dynamical breaking of the $\tilde\chi$-symmetry (D$\chi$SB) that is restored by the choice $\eta=\eta_{cr}$. It must be stressed that at variance with QCD the dynamically generated fermion mass itself is here O($\Lambda_s$), resulting in the squared mass of the pseudoscalar meson bound states to be O($\Lambda_s^2$) and hence comparable to that of other hadrons. 

2) {\it $\tilde\chi$-charge algebra closure} - A subtle question related to the unusual form of the mass terms that break the $\tilde\chi_L \times \tilde\chi_R$ WTIs~(\ref{CTLTI-RCR-NP}) and~(\ref{CTRTI-RCR-NP}) is whether (neglecting O($b^2$) terms) the algebra of $\tilde\chi$-charges closes. Although a rigorous analysis of this problem is beyond the scope of this paper, we can say that by suitably generalising standard chiral WTI arguments (see e.g.\ ref.~\cite{Luscher:1998pe}), one can positively answer the question. In fact, symmetry considerations imply that in products like $\tilde J_0^{L\,i} (x) \times (\bar Q_L \frac{\tau^j}{2} U Q_R - {\mbox{h.c.}})(0)$ no contact terms arise. 

3) {\it Naturalness} - The NP mass generation mechanism we have described in this work fulfils the 't Hooft naturalness requirement~\cite{THOOFT}, in the sense that the tuning of $\eta$ to its critical value has the effect of enlarging, even in the Nambu--Goldstone phase, the symmetries of the theory to include invariance under the chiral $\tilde\chi_L \times \tilde\chi_R$ transformations that only act on fermions.  

4) {\it NP mass counter-term subtraction} - An interesting question to ask is whether there is any field theoretically sound and ``natural'' way to subtract out the non-perturbatively generated mass term we have identified. The answer is negative. In fact, in order to eliminate all the NP $\tilde\chi_L \times \tilde\chi_R$ breaking effects from correlators, a counter-term, non-polynomial in the scalar fields and proportional to $\Lambda_s [\bar Q_L U Q_R + \bar Q_R U^\dagger Q_L]$, must be added to the fundamental Lagrangian~(\ref{SULL}). But the inclusion of such a counter-term non-polynomial in $\Phi$ would jeopardize the power-counting renormalizability of the basic model and also introduce in its UV-regulated action an hardly acceptable dependence on the phase (the counter-term makes sense only for $\hat\mu_\Phi^2 < 0$, i.e.\ $v\neq 0$) as well as on the RGI scale $\Lambda_s$.

5) {\it The magnitude of $v$} - If ideas of the kind developed in this paper are to be exploited to generate masses for fermions in alternative to the Higgs mechanism, one needs to assume that the vev of $\Phi$ satisfies the inequalities $\Lambda_s\ll v \ll b^{-1}$. The main reason is that, if instead $v \sim \Lambda_s$, one would be back to the situation where fermion masses are of the order of $\langle \Phi\rangle$, like in the Standard Model. Notice also that interestingly in the kinematical regime $\Lambda_s\ll v \ll b^{-1}$ the physics of the whole critical model at energies below $v$ turns out to be $v$-independent, because the $\sigma$ particle which has a square mass $m_\sigma^2\sim \hat\lambda v^2$ decouples. The condition $v\ll b^{-1}$  is needed to guarantee the independence of $\eta_{cr}$ on the value of $\hat\mu^2_\Phi$ 
(and its sign), thereby making unambiguous the step of $\tilde\chi_L\times\tilde\chi_R$-symmetry restoration, which is in turn essential to solve the naturalness problem. 

6) {\it The triviality issue of the scalar sector} - A question that deserves some discussion is the issue of the triviality of the scalar sector of the model~(\ref{SULL})~\footnote{We thank one of the anonymous referees for drawing our attention to this point.}. Triviality implies that the UV-cutoff can be made very large (compared to the renormalized scalar mass, $\hat\mu_\Phi$, and any other physical scale of the model), but may not be completely removed because the renormalized scalar quartic coupling, $\hat\lambda$, would approach zero as $b^{-1} \!\to\! \infty $ (at fixed values of the other renormalized parameters). This is most probably the case in the Wigner phase but, in view of the very peculiar NP effects we are advocating and the resulting effective interactions of NP origin between fermion and scalars, it is not at all clear whether this conclusion also holds in the Nambu--Goldstone phase, because an effective scalar quartic coupling 
O($\Lambda_s^2/v^2$) may survive as $b^{-1}\to \infty$. Moreover, it is not obvious 
at this stage whether the UV-cutoff should be finally removed or whether it might actually play a physical role as a very high energy scale where something else (say gravity) comes into play. 

As for the more practical question of the feasibility of lattice simulations of the model~(\ref{SULL}), the issue of triviality does not seem to pose any problem in numerical NP studies in view of the analyses of the Higgs model with standard lattice regularizations carried out e.g.\ in refs.~\cite{LuWe,BBHN,MoMu-book,Bock:1989gg,Bock:1990tv}). These investigations show that, in spite of triviality, for, say, $b \hat\mu_\Phi \sim 0.01$, one still has $\hat\lambda \sim {\mbox O}(1)$, implying that there exists a wide scaling region where cutoff effects are comfortably small with $\hat\lambda$ still significantly larger than zero.

7) {\it Masses, mixing and NP violation of universality} - The magnitude of 
the NP fermion masses generated by the mechanism we discuss in this paper is intrinsically dependent on the choice of the $\tilde\chi$-breaking terms in the basic Lagrangian ${\cal L}_{\rm toy}$, including the (for the moment not fully specified) details of the UV-regularization. In the toy model Lagrangian~(\ref{SULL}) we took, as an example, the $\tilde\chi$-breaking terms to be represented by ${\cal L}_{Wil}$.

In the framework of perturbation theory all $d>4$ terms would represent irrelevant details of the UV-completion of the ({\it critical}) model. But in the Nambu--Goldstone phase at the NP level, owing to the phenomenon of D$\tilde\chi$SB, all such ``irrelevant'' $\tilde\chi$-breaking operators are expected to produce physically ``relevant'', i.e.\ O($b^0\Lambda_s$), effects stemming from NP mixings among operators of unequal dimensionality (see sect.~\ref{sec:NPMIX}).

If this phenomenon occurs, it would provide the first (to our knowledge) example of NP universality breaking in a renormalizable gauge model. A far from trivial expectation like the one we have described needs of course to be {\em checked (possibly falsified)} by means of numerical Monte Carlo simulations of the ${\cal L}_{\rm toy}$ model in its Nambu--Goldstone phase~\footnote{For this purpose a specific lattice UV-regularization of the model must be adopted. If a regularization based on naive lattice fermions is chosen, in the interesting Nambu--Goldstone phase at $\eta_{cr}$ one has to face the presence of doubler modes with mass O($v$) already at the perturbative level~\cite{Bock:1989gg,Bock:1990tv}. Still, by taking $v \gg \Lambda_s$, it should be possible to check whether or not the fermion mode that in perturbation theory is massless receives a NP mass of order $\Lambda_s$.}. 

From a more phenomenological point of view a NP breaking of universality means that precise predictions about fermion masses become possible only when the details of the model at very high energy scales ($\sim b^{-1} \gg v$) are specified. Actually constraints on the structure of the UV-completion of the theory already appear when restoration of the the $\tilde\chi_L\times\tilde\chi_R$-symmetry in the presence of weak interactions is enforced~\cite{FRNEW}. Anyway we expect that in a realistic extension of the toy model~(\ref{SULL}) ratios of masses can be predicted with significantly smaller uncertainties than individual particle masses~\cite{FRNEW}.

\vspace{.3cm}
An example of how elementary fermion mass ratios can be understood and the interesting implications for the mass hierarchy problem are illustrated in next section where we consider an extension of ${\cal L}_{\rm toy}$ in which a new family of superstrongly interacting fermions is included.

\section{Strong meet superstrong interactions}
\label{sec:TECH}

In this section we want to examine a very interesting scenario for model-builders that occurs if besides ordinary quarks an extra family of fermions exists subjected to ordinary YM forces (whose gauge coupling we keep denoting by $g_s$) as well as to superstrong vector gauge interactions (with gauge coupling $g_T$). The superstrong force may be suggestively called techni-color, and the fermions subjected to it techni-fermions, with an eye to refs.~\cite{Weinberg:1979bn,Susskind:1978ms,Farhi:1980xs,Dimopoulos:1979es,Eichten:1979ah,Peskin:1997ez}, though our framework is very different from standard techni-color. 

In this system of coupled (asymptotically free) vector gauge interactions one can arrange things in such a way that the the modulus of the first coefficient of the superstrong $\beta$-function, $\beta^0_T$, is (appreciably) larger than that of the analogous coefficient of the YM interaction $\beta$-function, $\beta^0_{s}$. For instance, if one takes $N_{g}=3$ generations of ordinary (Dirac) quarks and one generation of (Dirac) techni-quarks, assuming $N_{c}=N_{T}=3$ for the color and techni-color gauge group and including weak isospin multiplicity, one gets $\beta^0_T/\beta^0_{s}=(11N_T-4N_c)/(11N_c-4N_g-4N_T)=7/3$.

Just like in the case of the model~(\ref{SULL}), we ought to include in the basic Lagrangian Wilson-like terms both for techni-fermions (with covariant derivatives depending on strong and superstrong gauge fields) and quarks, as well as the appropriate Yukawa terms. The $\Phi$ kinetic term and the scalar potential are like in~(\ref{SULL}). 

While under the exact $\chi_L \times \chi_R$ symmetries scalars, quarks and techni-fermions are simultaneously transformed, it is possible now to separately define transformations $\tilde\chi_L^q \times \tilde\chi_R^q$ acting only on quarks and transformations $\tilde\chi_L^{T} \times \tilde\chi_R^{T}$ acting only on techni-fermions. The critical model is hence defined by the requirement that the Yukawa terms for quarks and techni-fermions with coefficients $\eta_{cr}^q$ and $\eta_{cr}^{T}$, respectively, be such that in the Wigner phase of the model the WTIs of $\tilde\chi_L^q \times \tilde\chi_R^q$ and $\tilde\chi_L^{T} \times \tilde\chi_R^{T}$ are unbroken up to O($b^2$). In the Nambu--Goldstone phase, where $\langle \Phi \rangle = v > 0$, in analogy with the situation we discussed in sects.~\ref{sec:FINE1} and~\ref{sec:FINE2}, we expect dynamical spontaneous breaking of both $\tilde\chi_L^q \times \tilde\chi_R^q$ (driven by strong forces) and $\tilde\chi_L^{T} \times \tilde\chi_R^{T}$ (owing to superstrong interactions) symmetries.

\begin{figure}[htbp]
\vspace{-.5cm}
\centerline{\includegraphics[scale=0.70,angle=0]{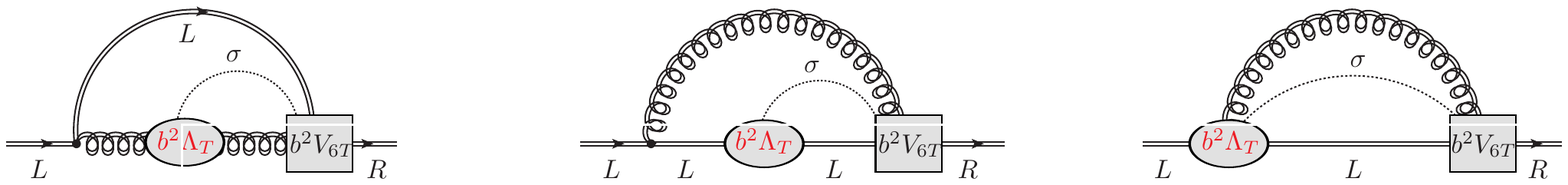}}
\vspace{-18.cm}
\caption{\small{Typical non-perturbative techni-fermion self-energy ``diagrams'', analogous to those in fig.~\ref{fig:LEAD-NPMIX-DIAG}, with the insertion of the techni-Wilson vertex, $b^2V_{6T}$. The grey blobs stand for the NP superstrong correction to the techni-gluon-techni-gluon-scalar, techni-fermion-techni-fermion-scalar and techni-fermion-techni-gluon-scalar vertex, respectively. The black dot (grey square box) represents the standard (techni-Wilson) vertex.}}
\label{fig:FIG7}
\end{figure}

Similarly to what we have conjectured it happens to $Q$-fields in the case of the model ${\cal L}_{\rm toy}$, here techni-fermions will acquire a non-perturbatively generated mass of the order $g_T^2\alpha_T\Lambda_{T}$ from ``diagrams'' similar to the ones in fig.~\ref{fig:LEAD-NPMIX-DIAG}, which we display in fig.~\ref{fig:FIG7}. In these figures double straight and curly lines represent techni-fermions and techni-gluons, respectively. As before, a dotted line represents a propagating $\sigma$ field. 
The grey blobs on the techni-gluon and techni-quark propagator stand for the non-perturbative contribution analogous to the one we have identified in sect.~\ref{sec:FINE2}, but here proportional to $b^2\,\Lambda_T\,\alpha_T$. The black dot and grey square box represent standard and techni-Wilson vertices, respectively.

\begin{figure}[htbp]
\vspace{-.8cm}
\centerline{\includegraphics[scale=0.5,angle=0]{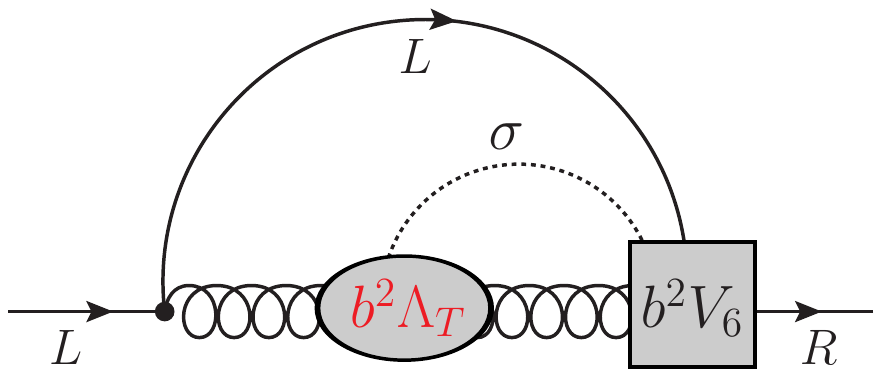}} 
\vspace{-11.5cm}
\caption{\small{A typical non-perturbative quark self-energy``diagram'' with the insertion of the standard Wilson vertex $b^2 V_{6}$. The grey blobs stand for the non-perturbative superstrong correction to the gluon-gluon-scalar vertex.}}
\label{fig:FIG8}
\end{figure}

Something quite interesting happens for ordinary quarks, because the mass contributions coming from the ``diagrams'' of fig.~\ref{fig:LEAD-NPMIX-DIAG} should now be replaced by those coming from ``diagrams'' like the one in fig.~\ref{fig:FIG8} where techni-fermions contribute to the NP correction of the gluon-gluon-scalar vertex. Terms of this kind are of order $g^2_s\alpha_s\Lambda_{T}$. 
Notice that to this order in $g_s^2$ the quark-quark-scalar vertex receives no analogous correction. As $\Lambda_{T}\gg \Lambda_{QCD}$, these self-energy contributions are much larger than the ones we have discussed in the previous sections, and are expected to completely dominate the effective value of the quark mass. 

We see that the leading contributions to quark and techni-fermion masses are thus both proportional to $\Lambda_T$, i.e.\ to the largest of the dynamically generated RGI scales~\footnote{Strictly speaking the notion of a hierarchy of RGI scales ($\Lambda_T \gg \Lambda_{QCD}$) is only valid to one-loop order in RG-improved perturbation theory. As soon as one goes to higher orders, the RG evolution equations of the various gauge couplings get coupled and only the RGI scale of the full theory has a meaning. In the situation of interest here this scale is to be identified with $\Lambda_T$.}, but multiplied by the fourth power of the coupling constant of the strongest among the vector gauge interactions the particle is subjected to. According to the considerations we have developed in the previous sections (see in particular the discussion in sect.~\ref{sec:DLMM} for the case of the critical ${\cal L}_{\rm toy}$ model) we get for fermion masses at the UV-cutoff scale to leading order in the gauge couplings the estimates 
\beqn
&&m_q^{dyn} = k_{LO}^{(q)} g_s^2 \alpha_s \Lambda_{T} \, , \qquad k_{LO}^{(q)} = {\mbox{O}}(1)  \label{MASSEQP}\, , \\
&&m_{T}^{dyn} =  k_{LO}^{(T)} g_T^2 \alpha_T \Lambda_{T} \, , \qquad k_{LO}^{(T)} = {\mbox{O}}(1)  \, .\label{MASSETP}
\eeqn
The interest of these formulae lies in the fact that they show that, in the coupled colour--techni-colour theory we have outlined, the quarks acquire an effective mass substantially smaller than the one of techni-fermions, because the mass of the former is scaled down by the fourth power of the ratio, $g_s/g_T$, of the two gauge couplings. 
For phenomenological considerations it is important to go beyond the leading order 
formulae above by using at least leading-log improved perturbative expressions
(written in terms of the appropriate renormalized couplings) and decide at what scale the effective fermion masses should be evaluated. 

The scope of possible phenomenological applications within the model scenario considered in this section is clearly limited not only by our ignorance of the radiative corrections to the diagrams in figs.~\ref{fig:FIG7} and~\ref{fig:FIG8}, but also by the as yet unrealistic matter content and the omission of electro-weak interactions. However, by making use of the concept of running effective fermion mass $m_Q^{dyn}(\mu)$ we introduced in Appendix~C (see eq.~(\ref{mQ-running})), we can roughly estimate the ratio $m_T^{dyn}/ m_q^{dyn} $ of techni-quark to quark masses at a convenient scale, denoted by $\mu_T$, where (in the scheme of choice) $\alpha_T(\mu_T) \sim 1/2$. 

The choice of the scale $\mu_T$ rather than $\Lambda_T$ itself (with $\alpha_T(\Lambda_T) = {\mbox{O}}(1)$) is due to the need of not completely loosing control of higher order corrections with respect to the RG-improved perturbative formulae we are going to use. On the other hand, by simple analogy between the assumed superstrong interactions and QCD, in the $\overline{MS}$ scheme we can expect the scale $\mu_T$ defined above to be only 2--3 times larger than $\Lambda_T$. Since ``techni-hadrons'' (gauge invariant bound states made out of valence techni-quarks) are expected to have a mass of the same order of magnitude as $\mu_T$, while the running of $m_q^{dyn}(\mu)$ from 
$\mu = \mu_T$ down to $\mu =  m_q^{dyn}$ is mild and well under control 
(at least for top or bottom quark), the estimate of $m_T^{dyn}(\mu_T)/ m_q^{dyn}(\mu_T) $ appears to be phenomenologically interesting.

Based on eq.~(\ref{mQ-running}) and noting $\tilde Z_m^{(q)}(\Lambda_T/\mu_T) = 1 + 
{{\mbox{O}}}(\alpha_s(\mu_T)) $ as well as $\tilde Z_m^{(T)}(\Lambda_T/\mu_T) = 1 + 
{{\mbox{O}}}(\alpha_T(\mu_T)) + {{\mbox{O}}}(\alpha_s(\mu_T)) $, beyond leading order 
in the gauge coupling(s) we can write~\footnote{To simplify formulae we use here 
the relation(s) $\alpha_{T,s}(\Lambda_T) = \alpha_{T,s}(\mu_T) [1 + {{\mbox{O}}}(\alpha_{T,s}(\mu_T))] $.}
\beqn
&&\hspace{-0.5cm}m_q^{dyn}(\mu_T) = k_{LO}^{(q)} g_s^2(\mu_T) \alpha_s(\mu_T) \Lambda_{T}
[ 1 + {{\mbox{O}}}(\alpha_s(\mu_T)) ]  \, ,  \label{MASSEQP-ao}\, , \\
&&\hspace{-0.5cm}m_{T}^{dyn}(\mu_T) =  k_{LO}^{(T)} g_T^2(\mu_T) \alpha_T(\mu_T) \Lambda_{T}
[ 1 + {{\mbox{O}}}(\alpha_T(\mu_T)) + {{\mbox{O}}}(\alpha_s(\mu_T)) ] \, . \label{MASSETP-ao}
\eeqn
As $\alpha_T(\mu_T) \gg \alpha_s(\mu_T)$ we get for the mass ratio
\beq
\frac{ m_{T}^{dyn}(\mu_T) }{ m_q^{dyn}(\mu_T) } \simeq 
\frac{ k_{LO}^{(T)} }{ k_{LO}^{(q)} } \frac{\alpha_T^2(\mu_T) }{\alpha_s^2(\mu_T) }
[ 1 + {{\mbox{O}}}(\alpha_T(\mu_T)) ] \, , \label{mTtoQ-1}
\eeq 
from which, assuming a similar pattern of $\tilde\chi$--breaking 
at the UV cutoff scale for techni-fermions and (the third generation of) quarks, 
which implies $k_{LO}^{(T)}/k_{LO}^{(q)} \simeq 1$, and inserting the values of
$\alpha_s(\mu_T)$  and $\alpha_T(\mu_T)$, we get
\beq
\frac{ m_{T}^{dyn}(\mu_T) }{ m_q^{dyn}(\mu_T) } \simeq
25 \times (1 \pm 0.5) \, . \label{mTtoQ-2}
\eeq
Identifying the quark flavour $q$ with the top 
(for reasons to be discussed in ref.~\cite{FRNEW})
and using the experimental value of its mass, we conclude that 
$m_{T}^{dyn}(\mu_T) \simeq 4 \times (1 \pm 0.5)$~TeV. In view of the discussion
above and in particular of eq.~(\ref{MASSETP-ao}) it also follows that the
the superstrong RGI scale $\Lambda_T$ is of the order of a few TeV's. To get a tighter prediction of the ratio~(\ref{mTtoQ-1}), as well as of the mass of the expected ``techni-hadrons'' and $\Lambda_T$, one needs to perform {\em ab initio} NP computations via Monte Carlo lattice simulations of the basic model. 

\section{Conclusions and outlook}
\label{sec:CONOUT}

In this paper we have discussed the implications of the possible existence in Wilson LQCD of a finite (up to logs) fermion mass contribution, dynamically generated as a result of the interplay between O($a$) chiral breaking effects left-over in the critical theory and the power divergency of loop integrals where a Wilson vertex is inserted. Effects of this kind turn out to contribute the critical mass a term of order $\alpha_s^2\Lambda_{QCD}$. Unfortunately, one cannot consider it as a {\it bona fide} quark mass because of the difficult ``fine tuning problem'' posed by the need of separating out the latter from the linearly diverging $1/a$ contribution that unavoidably goes with it.

We argue that this ``naturalness'' problem can be solved in an extension of QCD where a scalar field, coupled to a SU(2) doublet of fermions via a Yukawa interaction and a Wilson-like term, is introduced. We conjecture that, once in the Wigner phase of the model the Yukawa coupling has been tuned to a critical value where (up to negligible O(UV-cutoff$^{-2}$) effects) the scalar field decouples, the theory exhibits in its Nambu--Goldstone phase dynamically generated ``small'' fermion masses of the order $\alpha_s^2\Lambda_{s}$.      
   
The ``smallness'' of the dynamically generated fermion mass (as compared to the vev of the scalar field) is the consequence of the fact that at the critical Yukawa coupling ($\eta=\eta_{cr}$) the fermion chiral $\tilde\chi_L \times \tilde\chi_R$ transformations become (up to O(UV-cutoff$^{-2}$) corrections) a symmetry of the theory. In particular the cancellation of the ``large'' O($v$) quark mass term is guaranteed by the tuning of $\eta$ (see fig.~\ref{fig:VCANC}). The $\tilde\chi_L \times \tilde\chi_R$ charge algebra closes, even if in the Nambu--Goldstone phase the corresponding WTIs are broken by O($\alpha_s^2\Lambda_{s}$) mass terms of NP origin. 

The generation of such mass terms is triggered by the dynamical breaking of the recovered $\tilde\chi_L \times \tilde\chi_R$ symmetry. The precise magnitude of the NP mass depends on the details of Wilson-like terms present in the UV-regulated basic action for which at the moment we have no clue. It is conceivable, however, that fermion mass ratios are less sensitive to the UV-details of the $\tilde\chi_L \times \tilde\chi_R$-breaking terms than individual fermion masses. 

We thus see that, although the model is formally power-counting renormalizable, the non-perturbatively generated fermion masses appear to violate perturbative universality. This highly non-trivial conjecture is a natural conclusion of the arguments presented in this paper. It appears as a key point of the approach we propose. As such, it deserves in our opinion a dedicated numerical study via Monte Carlo simulations in order to confirm or falsify it.

If one accepts the kind of ``natural'' solution we have described in this paper for the ``fine tuning'' problem associated with the need of separating a ``large'' (perturbative) mass term from a ``small'' (NP) contribution, the peculiar gauge coupling dependence of the dynamically generated fermion mass could open the way to an interesting new approach to the mass hierarchy problem, according to which, schematically, the stronger is the strongest of the interactions a fermion is subjected to, the larger is its mass. In our opinion the NP mechanism for elementary particle mass generation we have presented in this work is much more ``natural'' than the situation one has if SM is taken as a Fundamental Theory. In fact, in the case of the SM even the order of magnitude of elementary fermion, weak gauge boson and Higgs masses is not understood, but rather merely fit to the experimental data. 

In our scenario, instead, although with the unavoidable limitations entailed by our ignorance of the UV model completion, at the price of introducing a new non-Abelian gauge interaction with a RGI-scale of few TeV's as well as new fermions coupled to both new and SM interactions with NP masses also in the TeV range, the origin of elementary particle masses is explained in terms of a common NP physical mechanism and their order of magnitude is parametrically understood.

Naturally, we cannot close this paper without commenting on the scalar resonance with mass of about 125~GeV recently discovered at LHC. In the scheme we are advocating here we would like to propose to interpret it as a two electro-weak bosons bound state, where the binding occurs through (yet to be discovered) superstrongly interacting fermions to which weak bosons are coupled due to weak interactions. Indeed a rough calculation of the $WW/ZZ$-binding energy~\cite{FRNEW} gives for it an estimate that is of the order of magnitude of the weak boson mass itself.

\vspace{0.5cm}

{\bf Acknowledgments -}  We wish to thank G.\ Martinelli, M.\ Testa, G.\ Veneziano and P.\ Weisz for numerous, illuminating  discussions. We also acknowledge useful comments from M.\ Bochicchio, V.\ Lubicz, N.\ Tantalo and C.\ Urbach. We thank P.\ Dimopoulos and D.\ Palao for a few valuable numerical checks in lattice QCD in the early stage of this work. 

\appendix 
\section{NP contributions to the fermion self-energy}  
\renewcommand{\thesection}{A} 
\label{sec:APPA} 

\subsection{Introduction}
In their seminal paper Banks--Casher~\cite{Banks:1979yr} conjectured that as a consequence of the phenomenon of S$\chi$SB, the eigenvalue density of the (Euclidean) Dirac operator in QCD does not vanish at $\lambda=0$, rather it behaves like 
\beq
\widehat\rho_{D}(0)=r_1\Lambda_{QCD}^3 \, ,
\label{RDOP}
\eeq
where by the symbol $\widehat .$ we mean averaging over gluons and sea quarks. The argument we develop in this Appendix is based on the idea of enriching/extending this assumption, by postulating a behaviour, at non-vanishing $\lambda$, of the kind 
\beq
\widehat \rho_{D}(\lambda)=r_1\Lambda_{QCD}^3+r_2\Lambda_{QCD}^2|\lambda|+r_3\Lambda_{QCD}|\lambda|^2+r_4|\lambda|^3+ \dots\, ,
\label{RDOPL}
\eeq
where, as we shall see, the term responsible for the emergence of a NP finite fermion mass is the third one, linear in $\Lambda_{QCD}$ and quadratic in $\lambda$, while the last term represents the kind of behaviour expected in perturbation theory (PT). Terms odd in $\Lambda_{QCD}$ are related to the phenomenon of S$\chi$SB~\cite{Smilga:1993in,Damgaard:1998xy,Osborn:1998qb}.

\begin{figure}[htbp]
\centerline{\includegraphics[scale=0.70,angle=0]{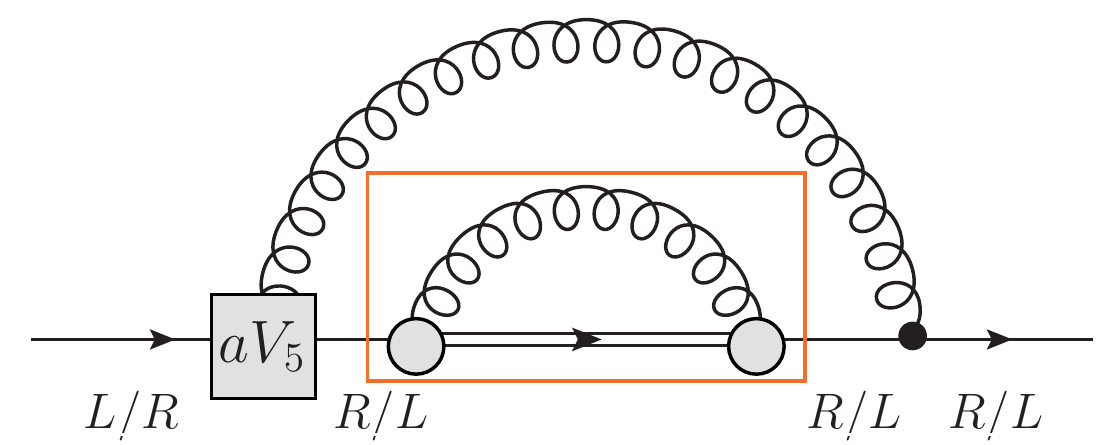}}
\caption{\small{A typical contribution to the fermion self-energy corresponding to the central term in fig,~\ref{fig:FIG6} of the paper. The square is a Wilson vertex. The continuum-like vertex is represented by a black dot. The pair of grey circles stands for any combination of the standard continuum-like and Wilson vertices. The double line means that the NP expression $\widehat \rho_{D}(\lambda)$ in eq.~(\ref{RDOPL}) is being taken in the spectral representation of the internal fermion propagator. The box encircles what in the text is called $S(x,y)|_{\rm{1loop}}$.}}
\label{fig:self1}
\end{figure}

In this Appendix we sketch the calculation of the fermion self-energy diagrams, one of which is drawn in fig.~\ref{fig:self1}. The calculation is (morally) based on the idea of expanding in PT the Schwinger--Dyson integral equations for propagators and vertices (described e.g.\ in~\cite{Alkofer:2000wg}) where we employ for the internal full fermion propagator a NP expression based on the form~(\ref{RDOPL}) of the eigenvalue density. 

Naturally this calculation cannot be carried out in full rigour/generality up to the end, otherwise we would have achieved the impossible goal of ``analytically'' computing the NP mass of an elementary particle. Our strategy will then be to work at lowest order in the counting of gauge couplings that in the diagram turn out to be evaluated at high momenta, while ideally summing sum over all soft gluon corrections giving rise to the NP modification of the internal fermion propagator. In order to avoid as much as possible uncontrolled approximations, we will try to reduce the necessary mathematical manipulations to general properties of spectral theory. 

\subsection{Lattice Dirac--Wilson operator - Spectral representation}
\label{sec:LATTWD}

To simplify calculations we take for the lattice Dirac--Wilson operator the expression 
\beq
D^{DW}=\slash D - \frac{ar}{2}\slash D \slash D \, .
\label{LIGDW}
\eeq
Without loss of generality as far as the chiral properties of $D^{DW}$ are concerned and only for the purpose of the argument/calculation presented here, we have chosen the particular Wilson term of eq.~(\ref{LIGDW}) so as to make $D^{DW}$ a normal operator. Owing to this property $D^{DW}$ can be diagonalised in an orthonormal basis. Furthermore, if we can solve the eigenvalue problem for $\slash D$, then obviously the one for $D^{DW}$ is also solved. Indeed, from 
\beq
\slash D_{\alpha\beta} F^s_\beta(x;\lambda) =i\lambda\, \epsilon_s F^s_\alpha(x;\lambda)\, ,
\label{EWD}
\eeq
one gets 
\beqn
&&D^{DW}_{\alpha\beta} F^s_\beta(x;\lambda)=i\tau(\lambda) F^s_\alpha(x;\lambda) \, ,\label{EDW1}\\
&&i\tau(\lambda)=i\lambda\epsilon_s +\frac{ar}{2}\lambda^2\, ,\label{EDW2}
\eeqn
where $\lambda$ is a non-negative real number and $\epsilon_s=+1$ for $s=1,2$ and $\epsilon_s=-1$ for $s=3,4$. 

Naturally eigenvalues and eigenfunctions depend on the gauge field appearing in $\slash D$ and spectral formulae hold for generic gauge field configurations. Averaging over the gauge configurations (with some gauge fixing), we gain translation and rotation invariance, and we write the ``averaged spectral formulae'' in the form
\beqn
\hspace{-.5cm}&&\delta^4(x-y)\delta_{\alpha\beta}=\sum_s\int d\lambda \int \frac{d\Omega_n}{2\pi^2} \widehat \rho_{D}(\lambda)\, \widehat F^s_\alpha (x;\lambda n)\widehat F_\beta^{s*}(y;\lambda n)\, ,\label{COMPL1}\\
\hspace{-.5cm}&&\widehat S^{DW}_{\alpha\beta}(x,y)=\sum_s\int d\lambda \int \frac{d\Omega_n}{2\pi^2} \frac{\widehat \rho_{D}(\lambda)}{i\lambda\epsilon_s+ar\lambda^2/2}\, \widehat F^s_\alpha (x;\lambda n) \widehat F_\beta^{s*}(y;\lambda n)\, .\label{COMPL2}
\eeqn
Rotation invariance of the gauge-averaged quark propagator, $\widehat S^{DW}$ is at the origin of the $\Omega_n$-integration in the two above equations.

\subsection{Computing the fermion propagator}
\label{sec:CSELF1}

For the reasons explained in the previous section we shall work with the fermionic action
\beq
{\cal S}_F=\int d^4x\,  \bar\psi\left(\slash D +\frac{ar}{2}\overleftarrow{\slash D} \slash D \right)\psi \, .
\label{FERMA}
\eeq
With an eye to the form of the Schwinger--Dyson equations, we see that to leading order in the gauge coupling the whole set of terms contributing to the NP correction to the fermion propagator (the block encircled by the rectangle in fig.~\ref{fig:self1}) can be compactly represented by the formula (Dirac indices are understood)
\beqn
\hspace{-1.5cm}&&S(x,y)\Big{|}_{\rm{1loop}}=\nn\\
\hspace{-1.5cm}&&=\int d^4x'\int d^4y' G_{\mu\nu}(x',y')\Big{\{}S(x,x') ig_s\Big{[}\gamma_\mu+\frac{ar}{2}\left({\overleftarrow{\slash D}}_{x'}\gamma_\mu-\gamma_\mu{\overrightarrow{\slash D}_{x'}}\right)\Big{]}\cdot\nn\\
\hspace{-1.5cm}&&\cdot {\widehat S}^{DW}(x',y') ig_s\Big{[}\gamma_\nu+\frac{ar}{2}\left({\overleftarrow{\slash D}}_{y'}\gamma_\nu-\gamma_\nu{\overrightarrow{\slash D}_{y'}}\right)\Big{]} S(y',y) \Big{\}}\, .
\label{GPFG1}
\eeqn
In eq.~(\ref{GPFG1}) $S(x,x')$ is the free fermion propagator, ${\widehat S}^{DW}(x',y')$ is given by eq.~(\ref{COMPL2}) and in the gauge we are using the free gluon propagator, $G_{\mu\nu}(x',y')$, reads 
\beq
G_{\mu\nu}(x',y')=\delta_{\mu\nu}\int \frac{d^4k}{(2\pi)^4}\frac{e^{ik(x'-y')}}{k^2} \, .
\label{GLUP}
\eeq
In fig.~\ref{fig:self1} the square is the Wilson vertex corresponding to the fermion action~(\ref{FERMA}). The continuum-like quark-gluon vertex is represented by a black dot. The double line means that the NP expression $\widehat \rho_{D}(\lambda)$ given by eq.~(\ref{RDOPL}) is being taken in the spectral representation of the internal fermion propagator. The pair of grey circles stands for any combination of the standard continuum-like and Wilson-like vertex over which we have to sum to get all the terms contributing to this order in $g_s^2$. Introducing the Fourier transform of the external free fermion propagator 
\beq
S(x,x')=\int \frac{d^4p}{(2\pi)^4} \frac{e^{ip(x-x')} }{\slash p} \, , 
\label{FERMP}
\eeq
and recalling eqs.~(\ref{COMPL2}) and~(\ref{EWD}), we get from~(\ref{GPFG1})
\beqn
\hspace{-.8cm}&&S(x,y)\Big{|}_{\rm{1loop}}\!\!=-g^2_s\!\!\int\! d^4x'\!\int\! d^4y'\! \int \!\frac{d^4p}{(2\pi)^4} \!\int\! \frac{d^4p'}{(2\pi)^4} \!\int \!\frac{d^4k}{(2\pi)^4} \, \frac{e^{ik(x'-y')}}{k^2} \sum_s \!\int \!d\lambda \!\int \!\frac{d\Omega_n}{2\pi^2} \nn\\
\hspace{-.8cm}&&\frac{e^{ip(x-x')}}{\slash p} \Big{[}1+\frac{ar}{2}\left(-i \slash p -i\lambda\epsilon_s \right)\Big{]}\frac{\widehat \rho_{D}(\lambda)}{i\lambda\epsilon_s+ar\lambda^2/2}\, \gamma_\mu \widehat F^s (x';\lambda n) \widehat F^{s*}(y';\lambda n)\gamma_\mu\cdot\nn\\
\hspace{-.8cm}&&\cdot \Big{[}1+\frac{ar}{2}\left(-i\lambda\epsilon_s-i{\slash p}\,\,'\right)\Big{]}\frac{e^{ip'(y'-y)} }{{\slash p}\,\,'}\, .
\label{GPFG2}
\eeqn
To proceed it is convenient to separate the chiral-breaking ($LR/RL$) and the chiral-preserving ($LL/RR$) part of the fermion propagator. The first correspond to terms with an overall even number of gamma-matrices and the second to terms with an odd number of gamma-matrices. In this counting, in view of eqs.~(\ref{EWD}), (\ref{COMPL1}) and~(\ref{COMPL2}), each factor $\epsilon_s$ should be considered as one gamma-matrix.

For the calculation of the fermion propagator to be consistent with the assumptions embodied in the eqs.~(\ref{DELTA2}) and~(\ref{ASYNT-FF-LQCD2}) of the paper we need first of all to check that to order $g_s^2$ no O($\Lambda_{QCD})$ mass terms get generated in its chiral-breaking part. 

\subsubsection{The chiral-breaking part of $S(x,y)\Big{|}_{\rm{1loop}}$}

We limit ourselves to considering in~(\ref{GPFG2}) only terms that do not contain
(in the numerator) explicit $\slash p$ and/or ${\slash p}\,\,'$ factors~\footnote{The neglected terms in fact give NP chiral breaking contributions to 
$S(p)|_{\rm{1loop}}^{LR,RL}$ of order $g_s^2a^2 p^2 \Lambda_{Q CD}$, the occurrence of which yields to 2-loop level a fermion mass term contribution of order $g_s^4  \Lambda_{QCD}$, i.e.\ of the same kind as the one we shall find in sect.~A.4 from the terms we retain.}. After the rewriting 
\beq
\frac{1}{i\lambda\epsilon_s+ar\lambda^2/2}=\frac{-i\lambda\epsilon_s+ar\lambda^2/2}{\lambda^2+(ar\lambda^2/2)^2}\, ,
\label{RAZ}
\eeq
one just discovers that summing all the terms that have an even number of gamma-matrices (counting, as we said, each $\epsilon_s$ factor as one gamma-matrix) yields a $\lambda$--independent integrand (except for the $\widehat \rho_{D}(\lambda)$ factor coming from the integration measure and the eigenfunctions).

Consequently one simply gets
\beqn
\hspace{-1.3cm}&&S(x,y)\Big{|}_{\rm{1loop}}^{LR,RL}\!\!=-g^2_s\frac{ar}{2}\!\!\int\! d^4x'\!\int\! d^4y'\! \int \!\frac{d^4p}{(2\pi)^4} \!\int\! \frac{d^4p'}{(2\pi)^4} \!\int \!\frac{d^4k}{(2\pi)^4} \, \frac{e^{ik(x'-y')}}{k^2} \nn\\
\hspace{-1.3cm}&&\frac{e^{ip(x-x')}}{\slash p}\sum_s \!\int \!d\lambda \!\int \!\frac{d\Omega_n}{2\pi^2}  \widehat \rho_{D}(\lambda)\, \gamma_\mu \widehat F^s (x';\lambda n) \widehat F^{s*}(y';\lambda n)\gamma_\mu\frac{e^{ip'(y'-y)} }{{\slash p}\,\,'}\, .
\label{GPFG3}
\eeqn
We see that in~(\ref{GPFG3}) the completeness relation~(\ref{COMPL1}) gets exactly reconstructed, so we obtain 
\beqn
\hspace{-1.1cm}&&S(x,y)\Big{|}_{\rm{1loop}}^{LR,RL}\!\!=-g^2_s\frac{ar}{2}\!\!\int\! d^4x'\!\int \!\frac{d^4p}{(2\pi)^4} \!\int\! \frac{d^4p'}{(2\pi)^4}  \frac{e^{ip(x-x')}}{\slash p}\!\int \!\frac{d^4k}{(2\pi)^4} \, \frac{1}{k^2} \frac{e^{ip'(x'-y)} }{{\slash p}\,\,'}=\nn\\
\hspace{-1.1cm}&&=-g^2_s\frac{ar}{2}\!\!\int \!\frac{d^4p}{(2\pi)^4}\frac{e^{ip(x-y)}}{p^2}\!\int\!\frac{d^4k}{(2\pi)^4} \, \frac{1}{k^2} \propto g^2_s\frac{r}{a}\!\int \!\frac{d^4p}{(2\pi)^4}\frac{e^{ip(x-y)}}{p^2}\, ,
\label{GPFG4}
\eeqn
i.e.\ the standard $1/a$ mass divergency. This calculation shows that to this order in $g^2_s$ no NP (finite) quark mass contribution gets generated.

\subsubsection{The chiral-preserving part of $S(x,y)\Big{|}_{\rm{1loop}}$}

Ignoring in~(\ref{GPFG2}) the terms containing (in the numerator) 
an explicit $a^2\slash p\,\,{\slash p}\,\,'$ 
factor~\footnote{The neglected terms actually give NP chiral 
preserving contributions to $S(x,y)|_{\rm{1loop}}^{LL,RR}$ at order $g_s^2$, 
the occurrence of which however yields to 2-loop level NP contributions to the
fermion mass of order $g_s^4  \Lambda_{QCD}$, i.e.\  of the same kind as the one we find in sect.~A.4 from the terms we retain.}, the calculation of the chiral-preserving part leads to the expression 
\beqn
\hspace{-1.3cm}&&S(x,y)\Big{|}_{\rm{1loop}}^{LL,RR}\!\!=g^2_s\!\!\int\! d^4x'\!\int\! d^4y'\! \int \!\frac{d^4p}{(2\pi)^4} \!\int\! \frac{d^4p'}{(2\pi)^4} \!\int \!\frac{d^4k}{(2\pi)^4} \, \frac{e^{ik(x'-y')}}{k^2} \nn\\
\hspace{-1.3cm}&&\frac{e^{ip(x-x')}}{\slash p}\sum_s \!\int \!d\lambda \!\int \!\frac{d\Omega_n}{2\pi^2}  \frac{\widehat \rho_{D}}{i\lambda\epsilon_s}\, \gamma_\mu \widehat F^s (x';\lambda n)  \widehat F^{s*}(y';\lambda n)\gamma_\mu\frac{e^{ip'(y'-y)} }{{\slash p}\,\,'} \, .
\label{GPFG5}
\eeqn
Spectral properties allow us to rewrite eq.~(\ref{GPFG5}) in the compact form
\beqn
\hspace{-1.3cm}&&S(x,y)\Big{|}_{\rm{1loop}}^{LL,RR}\!\!=g^2_s\!\int\! d^4x'\!\int\! d^4y'\! \int \!\frac{d^4p}{(2\pi)^4} \!\int\! \frac{d^4p'}{(2\pi)^4} \!\int \!\frac{d^4k}{(2\pi)^4} \, \nn\\
\hspace{-1.3cm}&& \frac{e^{ik(x'-y')}}{k^2}\frac{e^{ip(x-x')}}{\slash p}(\widehat{\slash D})^{-1}(x',y')\frac{e^{ip'(y'-y)} }{{\slash p}\,\,'} \, .
\label{GPFG6}
\eeqn
Introducing the Fourier transform
\beq
(\widehat{\slash D})^{-1}(x',y')=\int \frac{d^4q}{(2\pi)^4}(\widehat{\slash D})^{-1}(q)\, e^{iq(x'-y')}\, ,
\label{FTD}
\eeq
eq.~(\ref{GPFG6}) becomes 
\beqn
\hspace{-1.3cm}&&S(x,y)\Big{|}_{\rm{1loop}}^{LL,RR}\!\!=g^2_s\!\!\int \!\frac{d^4p}{(2\pi)^4} 
\int \!\frac{d^4q}{(2\pi)^4} \,   
\frac{1}{\slash p}\, \frac{e^{ip(x-y)}}{(p-q)^2} \, 
(\widehat{\slash D})^{-1}(q)\frac{1}{{\slash p}} \, .
\label{GPF7}
\eeqn
Since no analytic NP expression of $(\widehat{\slash D})^{-1}(q)$ is obviously available, we have to make recourse to some kind of approximations. The latter are better introduced in the expression~(\ref{GPFG5}) by stipulating that the eigenfunctions of the averaged Dirac operator are the free ones and that $\widehat \rho_D$ is given by eq.~(\ref{RDOPL}). Thus under the replacement  
\beq
\sum_s \widehat F^s_\alpha (x';\lambda n)\epsilon_s  \widehat F^{s*}_\beta (y';\lambda n)  \to \frac{e^{i\lambda  n(x'-y')}}{(2\pi)^4}(\slash n)_{\alpha\beta} \, ,
\label{FTH}
\eeq
we obtain (up to irrelevant multiplicative constant factors)
\beqn \hspace{-1.5cm}&&S(x,y)\Big{|}_{\rm{1loop}}^{LL,RR}\!\!\propto g^2_s\!\!\int \!\frac{d^4p}{(2\pi)^4}\, \frac{e^{ip(x-y)}}{\slash p} \int \!d\lambda \!\int \!\frac{d\Omega_n}{2\pi^2}  \frac{\widehat \rho_{D}}{\lambda} \frac{\slash n}{(p-\lambda n)^2}
 \frac{1}{{\slash p}} \, . \label{GPF8_BIS}\eeqn
We evaluate below the NP contribution to the chiral preserving part of the
fermion propagator coming from the 1-loop integral in the r.h.s.\ of eq.~(\ref{GPF8_BIS})
in order to check that it takes just the form we conjectured in sect.~2.2 relying 
on Symanzik expansion arguments, see eqs.~(\ref{DELTA2}) and~(\ref{ASYNT-FF-LQCD2}). Setting $q_\mu = \lambda n_\mu$, $q = \sqrt{ q_\mu q_\mu }$ and focusing (symbol $\Rightarrow$) on the contribution of the piece $r_3 \Lambda_{QCD} \lambda^2$ in $\widehat \rho_{D}(\lambda)$ (see eq.~(\ref{RDOPL})) we obtain
\beqn \hspace{-1.5cm}&&
g_s^2 \int \!d\lambda \!\int \!\frac{d\Omega_n}{2\pi^2}  \frac{\widehat \rho_{D}}{\lambda} \frac{\slash n}{(p-\lambda n)^2 + \epsilon_{IR}^2} \Rightarrow
g_s^2 \int \!d q \!\int \!\frac{d\Omega_q}{2\pi^2} \frac{r_3 \Lambda_{QCD} \, \slash q}{(q - p)^2 + \epsilon_{IR}^2} 
\, . \label{evalGPF8_A}\eeqn
With respect to the the r.h.s.\ of eq.~(\ref{GPF8_BIS}) here an IR cutoff $\epsilon_{IR} \propto \Lambda_{QCD}$ has been inserted in the gluon propagator factor $1/[(q - p)^2 + \epsilon_{IR}^2]$. An IR cutoff term of this type is actually expected to be generated by higher order radiative corrections to the fixed gauge gluon propagator and will be useful in the small--$p^2$ expansion below. In order to make contact with eq.~(\ref{DELTA2}), we in fact perform in eq.~(\ref{evalGPF8_A}) a Taylor expansion around $p=0$ and find
\beqn \hspace{-1.2cm}&& 
g_s^2 \int \!d\lambda \!\int \!\frac{d\Omega_n}{2\pi^2}  \frac{\widehat \rho_{D}}{\lambda} \frac{\slash n}{(p-\lambda n)^2 + \epsilon_{IR}^2} \Rightarrow
g_s^2 r_3 \, \slash p \, \frac{\Lambda_{QCD}}{\epsilon_{IR}}  
\frac{\pi}{4}  [1 + {\mbox{O}}(a^2p^2,a^2\epsilon_{IR}^2)] +
\nonumber \\
\hspace{-1.2cm}&& - g_s^2 r_3 \, \slash p \, \Lambda_{QCD} \frac{a}{2}  
[1 + {\mbox{O}}(a^2p^2,a^2\epsilon_{IR}^2)]
\, , \label{evalGPF8_B}\eeqn
where the first term in the r.h.s.\ is a (UV-finite) part of the standard fermion propagator renormalization and the second one is the NP contribution. 

\subsection{Fermion self-energy}

The quantity $S(x,y)|_{\rm{1loop}}^{LL,RR}$ we have computed above is the internal loop in the rectangular box within the self-energy diagram of fig.~\ref{fig:self1}. For the computation of the contribution to the fermion mass we can set the external quark momentum to zero. Up to irrelevant (for these considerations) constant factors, introducing the chiral-preserving piece of the propagator we have just evaluated in the diagram of fig.~\ref{fig:self1}, we see that, focusing on the contribution coming from the term proportional to $\Lambda_{QCD}$ in $\widehat \rho_{D}$, we get the 2-loop integral expression 
\beq
m_q^{dyn} 
\propto a r_3 \, g^4_s \Lambda_{QCD} \int \!\frac{d^4p}{(2\pi)^4} \int \!\frac{d^4q}{(2\pi)^4} \, \frac{1}{p^2} \frac{p_\mu }{\slash p}\frac{\slash q}{q} \frac{1}{q^2} \, \frac{1}{(p-q)^2} \, \frac{ \gamma_\mu}{{\slash p}}\, ,
\label{DMQ}
\eeq
where we called $p$ the momentum of the outer loop (remind also $q=\lambda n$). 
The double integral in eq,~(\ref{DMQ}) is IR finite and diverges linearly in the UV. This $1/a$ divergence compensates the explicit multiplicative $a$ factor, 
leaving behind a finite contribution that is parametrically of the announced form
\beq
m_q^{dyn}  = {\mbox{O}}(g^4_s \Lambda_{QCD})\, .
\label{MQ}
\eeq

\appendix 
\renewcommand{\thesection}{B} 
\section*{Appendix B - The mixing pattern of operators~(\ref{HDOP-WIL}) and~(\ref{HDOP-WIR})}
\label{sec:APPB} 

In this Appendix we want to prove that the operator $O_{6}^{L\,i}$ and $O_{6}^{R\,i}$ defined in eqs.~(\ref{HDOP-WIL}) and~(\ref{HDOP-WIR}) can only mix with the $d<6$ operators appearing in the r.h.s.\ of eqs.~(\ref{O6L-MIX}) and~(\ref{O6R-MIX}), respectively. Given the exact parity symmetry of the toy-model Lagrangian~(\ref{SULL}), 
we can limit the discussion to, say, the operator $O_{6}^{L\,i}$. 

The operators $\partial_\mu\tilde J_\mu^{L\,i}$ and $[\bar Q_L\frac{\tau^i}{2}\Phi Q_R-{\mbox{h.c.}}]$ are easily seen to be the only (gauge invariant) $d<6$ quark bilinears enjoying the same properties as $O_{6}^{L\,i}$ under $\chi_L\times\chi_R$ and discrete $C$, $P$, $T$ and flavour symmetries. Thus we are only left with the task of excluding the $d<6$ operators constructed in terms of $\Phi$-fields in the following list
\beqn
&& O_2^{L\, i}={\rm tr}\Big{[}\Phi^\dagger\frac{\tau^i}{2}\Phi\Big{]}\label{P1}\\
&& O_{4,1}^{L\, i}={\rm tr}\Big{[}\Phi^\dagger\Phi\Big{]}{\rm tr}\Big{[}\Phi^\dagger\frac{\tau^i}{2}\Phi\Big{]}\label{P2}\\
&& O_{4,2}^{L\, i}={\rm tr}\Big{[}\partial_\mu\Phi^\dagger\frac{\tau^i}{2}\partial_\mu\Phi\Big{]} \label{P3}\\
&& O_{4,3}^{L\, i}={\rm tr}\Big{[}\Phi^\dagger\frac{\tau^i}{2}[\overleftarrow\partial\,^2+\partial^2]\Phi\Big{]}\label{P4}
\eeqn
This can be done on the basis of the $C$, $P$ and $F_2$ discrete transformations that are exact symmetries of ${\cal L}_{\rm toy}$. For completeness we recall here their definition  
\begin{equation}    
\hspace{-.9cm}\!\!P : \left \{\begin{array}{lll}
&\!\!\!\Phi(x)\rightarrow \Phi^\dagger (x_P)\, ,  & x_P\equiv(-\vec x, x_0)  \\
&\!\!\!Q(x)\rightarrow\gamma_0 Q(x_P) \, , & \bar Q(x)\rightarrow\bar Q(x_P)\gamma_0 \\     
&\!\!\!A_k(x)\rightarrow -A_k(x_P)\, ,  & A_0(x)\rightarrow A_0(x_P)    
\end{array}\right . \label{PTR} \end{equation} 
\begin{equation}    
C : \left \{\begin{array}{lll}
&\!\!\!\Phi(x)\rightarrow \Phi^T (x) \\
&\!\!\!Q(x)\rightarrow i\gamma_0\gamma_2 \bar Q^T(x) \, , & \bar Q(x)\rightarrow -Q^T(x)i\gamma_0\gamma_2 \\     
&\!\!\!A_\mu(x)\rightarrow -A_\mu^\star(x)  
\end{array}\right . \label{CTR1} \end{equation} 
\begin{equation}    
\hspace{-.9cm}\!\!\!\!F_2 : \left \{\begin{array}{lll}
&\!\!\!\Phi(x)\rightarrow \tau^2\Phi(x)\tau^2 \\
&\!\!\!Q(x)\rightarrow i\tau^2 Q(x) \, , & \bar Q(x)\rightarrow -i \bar Q(x)\tau^2      
\end{array}\right . \label{CTR2} \end{equation}
Looking at the way the various operators we are considering (i.e.\ $O_{6}^{L,i}$ and those listed in equations from~(\ref{P1}) to~(\ref{P4})) transform under $CP$ and $CPF_2$, we can construct Table~\ref{tab:tab1}.
\begin{table}[hbt]
  \centering
  \begin{tabular}{|c|c|c|c|}
  \hline
  Operator & $CP\, ,\, i=1,3$ & $CP\, ,\, i=2$ & $CPF_2$\\
    \hline\hline
 $O_2^{L\,i}$ & even & odd & odd\\
 \hline
  $O_{4,1}^{L\,i}$ & even & odd & odd\\
 \hline
  $O_{4,2}^{L\,i}$ & even & odd & odd\\ 
  \hline
 $O_{4,3}^{L\,i}$ & even & odd & odd\\
    \hline\hline
   $O_{6}^{L\,i}$ & odd & even & even\\
     \hline\hline  
  \end{tabular}
 \caption[tab:results]{\small{The parities of the operators $O_{6}^{L,i}$ and of those in equations from~(\ref{P1}) to~(\ref{P4}) under the discrete transformations $CP$ and $CPF_2$.}}
  \label{tab:tab1}
\end{table}
We see from Table~\ref{tab:tab1} that the operators from~(\ref{P1}) to~(\ref{P4}) have $CPF_2$ transformations properties opposite to that of $O_{6}^{L,i}$, so they cannot appear in the r.h.s.\ of eq.~(\ref{O6L-MIX})~\footnote{As it is usually done for isospin with $G$-parity, the transformation $F_2$ is introduced here to compensate for the different $CP$ properties of different weak isospin components.}.

We conclude this Appendix by recalling that, unlike the operator (\ref{P4}), the combination 
\beq
\partial_\mu {\rm tr}\Big{[}\Phi^\dagger\frac{\tau^i}{2}(\partial_\mu\Phi)-(\partial_\mu\Phi^\dagger) \frac{\tau^i}{2}\Phi\Big{]}={\rm tr}\Big{[}\Phi^\dagger\frac{\tau^i}{2}(\partial^2\Phi)-(\Phi^\dagger\overleftarrow\partial\,^2) \frac{\tau^i}{2}\Phi\Big{]}\, ,\label{P5}
\eeq 
because of the minus sign between the two bits of the operator, is even under $CPF_2$, just like $O_{6}^{L,i}$. As noted in sect.~\ref{sec:MSE-PT}, the latter can be eliminated in the mixing in favour of $\partial_\mu\tilde J_\mu^{L\,i}$ owing to the conservation of the $\chi_L$ current, $\partial_\mu J_\mu^{L\,i}=0$ (see the expressions of $J_\mu^{L\,i}$ and $\tilde J_\mu^{L\,i}$ in eqs.~(\ref{CCL}) and~(\ref{JCLT}), respectively).

\appendix 
\renewcommand{\thesection}{C} 
\section*{Appendix C - The running of NP masses} 
\label{sec:APPC} 

A crucial issue for the interpretation of $C_1\Lambda_s$ as a quark mass is the behaviour of the coefficient $C_1$ as a function of $\log b \Lambda_s$. Recalling the lowest order expression of $C_1$ given in eq.~(\ref{C1}), the WTIs~(\ref{CTLTI-RCR-NP}) and~(\ref{CTRTI-RCR-NP}) to the same order in $g_s^2$ can be cast in the form 
\beqn
\hspace{-1.4cm}&&\partial_\mu \langle Z_{\tilde J}\tilde J^{L\, i}_\mu(x) \,\hat O(0)\rangle\Big{|}_{cr}^{\rm LO} = \langle \tilde\Delta_{L}^i\hat O(0)
\rangle\Big{|}_{cr}^{\rm LO} \delta(x) + \nonumber \\
\hspace{-1.4cm}&&\qquad + k_{LO}\,g_s^2 \, \alpha_s \, \Lambda_s \, \langle
\,\Sigma^i_L(x)\,\hat O(0)\,\rangle\Big{|}_{cr}^{\rm LO} +{\mbox O}(b^2)\, ,\label{WTIRGIL-LO}\\
\hspace{-1.4cm}&&\partial_\mu \langle Z_{\tilde J}\tilde J^{R\, i}_\mu(x) \,\hat O(0)\rangle\Big{|}_{cr}^{\rm LO}  = \langle \tilde\Delta_{R}^i\hat O(0)
\rangle\Big{|}_{cr}^{\rm LO} \delta(x) + \nonumber \\
\hspace{-1.4cm}&&\qquad +k_{\rm LO}\, g_s^2 \, \alpha_s\, \Lambda_s \, \langle
\,\Sigma^i_R(x) \,\hat O(0)
\, \rangle\Big{|}_{cr}^{\rm LO} +{\mbox O}(b^2) \, ,\label{WTIRGIR-Lo}
\eeqn
where for short we have defined the local operators (also here for simplicity we ignore the possible appearance of the factor ${\cal P}$ we mentioned in sect.~\ref{sec:INU})
\beqn
&&\Sigma^i_L(x)= [\bar Q_L \frac{\tau^i}{2} U Q_R - {\mbox{h.c.}}\,](x) \, ,\label{SIGL}\\
&&\Sigma^i_R(x)= [\bar Q_R \frac{\tau^i}{2} U^\dagger Q_L - {\mbox{h.c.}}\,](x) \, .\label{SIGR}
\eeqn
Consistently with the renormalizability of our toy-model and the general arguments of ref.~\cite{Testa:1998ez}, we expect higher order radiative corrections to provide the correct RG evolution of all the quantities above. In particular the RGI of the l.h.s.\ of the WTIs~(\ref{CTLTI-RCR-NP}) and~(\ref{CTRTI-RCR-NP}) entails, as discussed in sect.~\ref{sec:NPMIX}, the same property for both the r.h.s.\ contributions. In the full theory the (NP terms in the) WTIs~(\ref{CTLTI-RCR-NP}) and~(\ref{CTRTI-RCR-NP}) must hence take the form
\beqn
\hspace{-1.4cm}&&\partial_\mu \langle Z_{\tilde J}\tilde J^{L\, i}_\mu(x) \,\hat O(0)\rangle\Big{|}_{cr} = \langle \tilde\Delta_{L}^i\hat O(0)
\rangle\Big{|}_{cr} \delta(x) + \nonumber \\
\hspace{-1.4cm}&&\qquad + k_{\rm LO}\,g_s^2(b^{-1}) \tilde Z_m(b\Lambda_s) \, \alpha_s(\Lambda_s)\, \Lambda_s \, \langle
\,\Sigma^i_L(x)\,\hat O(0)\,\rangle\Big{|}_{cr} +{\mbox O}(b^2)\, ,\label{WTIRGIL}\\
\hspace{-1.4cm}&&\partial_\mu \langle Z_{\tilde J}\tilde J^{R\, i}_\mu(x) \,\hat O(0)\rangle\Big{|}_{cr}  = \langle \tilde\Delta_{R}^i\hat O(0)
\rangle\Big{|}_{cr} \delta(x) + \nonumber \\
\hspace{-1.4cm}&&\qquad +k_{\rm LO}\, g_s^2(b^{-1}) \tilde Z_m(b\Lambda_s) \, \alpha_s(\Lambda_s) \, \Lambda_s \, \langle
\,\Sigma^i_R(x) \,\hat O(0)
\, \rangle\Big{|}_{cr}+{\mbox O}(b^2) \, ,\label{WTIRGIR}
\eeqn
where the dimensionless quantity $\tilde Z_m(b\Lambda_s)$ incorporates all the radiative correction effects in the NP fermion mass terms and admits the perturbative expansion
\beq
\tilde Z_m(b\Lambda_s)=1+g_s^2 (b^{-1})\, (\tilde \gamma_m\, \log b\Lambda_s+\tilde c_m ) \, + \, \ldots\,\, .
\eeq
Note in eqs.~(\ref{WTIRGIL})--(\ref{WTIRGIR}) the specification of the scale in the gauge coupling factors $g_s^2(b^{-1})$ and $\alpha_s(\Lambda_s)$. On the one hand this somewhat arbitrary choice of scales entails no loss of generality as it is actually part of our definition of $\tilde Z_m(b\Lambda_s)$. On the other here our rationale for this choice is simply that in the mass generation mechanism of sects.~\ref{sec:NGP}--\ref{sec:DLMM} we conjectured the occurrence of O($b^2 \Lambda_s \alpha_s)$ NP vertex corrections (for which higher order radiative effects are likely to yield $\alpha_s \to \alpha_s(\Lambda_s)$) that become non-irrelevant when combined with the UV power-divergent loop effect of relative order $g_s^2$ (for which we assume $g_s^2 \to g_s^2(b^{-1})$) .

By comparing eqs.~(\ref{WTIRGIL}) and~(\ref{WTIRGIR}) with eqs.~(\ref{CTLTI-RCR-NP}) and~(\ref{CTRTI-RCR-NP}) we see that
\beq
C_1 \Lambda_s = k_{\rm LO}\, g_s^2(b^{-1}) \tilde Z_m(b\Lambda_s) \, \alpha_s(\Lambda_s) \Lambda_s
\equiv m_Q^{dyn}(b^{-1}) \label{mQ-UVscale}
\eeq
represents indeed the fermion mass at the UV-cutoff scale $b^{-1}$, because it enters the $\tilde\chi_L \times \tilde\chi_R$ WTIs as the coefficient of the density $\Sigma^i_{L/R}$ at the UV-cutoff scale. 

This interpretation of the NP fermion mass terms in eqs.~(\ref{WTIRGIL}) and~(\ref{WTIRGIR}) and their (necessary) RG invariance imply
\beq
m_Q^{dyn}(b^{-1})\,\Sigma^i_{L/R}=m_Q^{dyn}(\mu)\,\hat \Sigma^i_{L/R}(\mu)\, ,
\label{MRGI}\eeq
where we have introduced the running NP fermion mass
\beq
 m_Q^{dyn}(\mu) = k_{\rm LO}\, g_s^2(\mu) \tilde Z_m(\Lambda_s/\mu)
 \, \alpha_s(\Lambda_s) \Lambda_s  \label{mQ-running}
\eeq
as well as the renormalized densities 
\beq
\hat \Sigma^i_{L/R}(\mu) = Z_\Sigma(b\mu) \Sigma^i_{L/R}(b^{-1})\, .   \label{Sigma-REN}
\eeq
Eq.~(\ref{MRGI}) shows that the dynamically generated fermion mass $m_Q^{dyn}$ {\em can} indeed be {\em interpreted} as a running mass and provides the desired RG equation for it. 

If we choose for the renormalization scale $\mu$ the value $\Lambda_s$, which is of phenomenological interest as it represents the natural NP mass scale of the model, we get 
\beq
 m_Q^{dyn}(\Lambda_s) = k_{\rm LO}\, g_s^2(\Lambda_s) \tilde Z_m(1)
 \, \alpha_s(\Lambda_s) \Lambda_s. \label{RGI}
\eeq 
We end by noting that RGI of NP mass terms in the WTIs~(\ref{WTIRGIL}) and~(\ref{WTIRGIR}) 
entails the relations
\beq
Z_{\Sigma} (b\mu) = \frac{g^2_s(b^{-1})\tilde Z_m(b\Lambda_s)}{g^2_s(\mu)\tilde Z_m(\Lambda_s/\mu)}\, .
\label{ZS-mu} 
\eeq
and
\beq
Z_{\Sigma} (b\Lambda_s) =\frac{g^2_s(b^{-1})\tilde Z_m(b\Lambda_s)}{g^2_s(\Lambda_s)\tilde Z_m(1)}\, .
\label{ZS-Lambda}
\eeq
where $g_s(\mu)$ is the renormalized gauge coupling at the $\mu$-scale. To first order in $g_s^2$ from the
relations above (or equivalently to O($g_s^6$) in eq.~(\ref{MRGI})), we have
\beq
-2\beta_0 + \tilde\gamma_m-\gamma_{\Sigma}=0 \, , 
\eeq
where we assumed the perturbative expansions
\beqn
\hspace{-1.5cm}&&Z_{\Sigma}(b\mu)=1+g_s^2 (b^{-1})\, (\gamma_\Sigma\,\log b\mu+c_\Sigma) \, +\ldots \, , \label{ZEXP}\\
\hspace{-1.5cm}&&\nn\\
\hspace{-1.5cm}&&g_s^2(\mu)=g_s^2(b^{-1})[1+g_s^2(b^{-1})\, (2\beta_0\,\log{b\mu}+c_\beta)\, +\ldots \,] \, .
\label{GRUN}
\eeqn

\end{document}